\documentclass{article}

\PassOptionsToPackage{numbers,compress}{natbib}
\usepackage[preprint]{neurips_2026}

\usepackage[utf8]{inputenc} 
\usepackage[T1]{fontenc}    
\usepackage{hyperref}       
\usepackage{url}            
\usepackage{booktabs}       
\usepackage{amsfonts}       
\usepackage{nicefrac}       
\usepackage{microtype}      
\usepackage{xcolor}         
\usepackage{listings}
\usepackage{graphicx}
\usepackage{amsmath}
\usepackage{array}
\usepackage{multirow}
\usepackage{rotating}
\newcolumntype{L}[1]{>{\raggedright\arraybackslash}p{#1}}

\makeatletter
\@ifundefined{nolinenumbers}{%
  \newenvironment{nolinenumbers}{}{}%
}{}
\makeatother

\title{When Does Hierarchy Help? Benchmarking Agent Coordination in Event-Driven Industrial Scheduling}

\lstdefinestyle{promptblock}{
    basicstyle=\ttfamily\footnotesize,
    frame=single,
    breaklines=true,
    breakatwhitespace=false,
    columns=fullflexible,
    keepspaces=true,
    showstringspaces=false,
    tabsize=2,
    xleftmargin=1em,
    xrightmargin=1em,
    framexleftmargin=0.8em,
    framexrightmargin=0.8em,
    backgroundcolor=\color{gray!5},
    rulecolor=\color{gray!50},
    framerule=0.4pt,
    aboveskip=0.8em,
    belowskip=0.8em
}
\author{%
  Ziqi Wang, Yuhao Yang\thanks{Project lead.}, Zhiwei Ling, Wenzhuo Qian, Hailiang Zhao\thanks{Corresponding author. Email: hliangzhao@zju.edu.cn} \\
  Zhejiang University \\
}

\begin{document}

\maketitle

\begin{abstract}
  Recent advances in agent and multi-agent systems have shown strong performance on tool use, reasoning, and collaborative tasks. However, existing benchmarks mostly evaluate task completion in weakly coupled environments, and provide limited support for studying coordination in shared, dynamically evolving systems with hierarchy and coupled constraints. This leaves an important question underexplored: when do different coordination paradigms succeed or fail? We introduce Distributed Event-driven Scheduling Benchmark (DESBench), a benchmark for evaluating agent coordination in hierarchical event-driven scheduling. Built on a shared discrete-event driven environment in industrial scheduling, our benchmark captures multi-timescale decision making, partial observability, and dynamically coupled constraints. We define tasks and metrics that evaluate effectiveness, constraint alignment, coordination efficiency, and robustness, and focus on four representative coordination paradigms: centralized, hierarchical, heterarchical, and holonic. These paradigms correspond to distinct mechanisms of information flow, decision authority, and conflict resolution. Our controlled evaluations reveal clear coordination trade-offs: centralized coordination is robust and communication-efficient but scales poorly with difficulty; hierarchical coordination improves efficiency through decomposition but suffers from cross-level misalignment; heterarchical coordination is flexible but communication-heavy; and holonic coordination satisfies constraints well but loses global robustness. These findings demonstrate that coordination design fundamentally shapes agent system behavior in complex environments, revealing structural trade‑offs that cannot be captured by outcome metrics alone and underscoring the imperative for more adaptive, principled, and dynamic coordination mechanisms in future MAS research. DESBench is available at: https://github.com/Mr-lander/DESBenchmark.

  
\end{abstract}

\section{Introduction}
The rapid progress of agent and multi-agent systems (MAS) has expanded their use beyond single-step reasoning and tool invocation toward more complex forms of coordination and collaborative problem solving \cite{deng2025agentic}. A growing body of work explores how multiple agents can coordinate to solve complex problems, often leveraging large language models (LLM) or learned policies \cite{li2024survey}. However, despite these advances, current evaluation benchmarks remain largely focused on task completion in weakly coupled environments, where interactions are short-horizon, state transitions are loosely coupled, and coordination plays a limited role in shaping long-term system behavior. In contrast, real‑world decision‑making often takes place in shared, dynamically changing environments where interaction through a common system state yields delayed and widespread effects \cite{wagenmaker2024overcoming,rohbeck2025modeling}. Industrial scheduling problems provide an excellent testing ground for these scenarios, where coordination is not merely a mechanism for dividing work, but a central factor that determines system performance, constraint satisfaction, and robustness. Existing benchmarks are insufficient for studying coordination for three key reasons:
\begin{itemize}
    \item Traditional scheduling and optimization benchmarks primarily evaluate solution quality under implicit centralized or fixed control assumptions, providing limited insight into how alternative coordination structures influence system behavior.
    \item Many existing agent benchmarks lack a fully shared and dynamically evolving environment, and therefore only partially capture coordination settings where actions influence future dynamics through resource coupling and delayed feedback.
    \item Current environments rarely integrate hierarchical organization with dynamically coupled constraints in a unified setting, making it difficult to systematically analyze how different coordination paradigms operate under multi-level decision processes.
\end{itemize}

As a result, a fundamental question remains underexplored: \textit{When do different coordination paradigms succeed or fail in complex and dynamic environments?} In the spirit of Aristotle's insight that \textit{the whole is greater than the sum of its parts}, coordination mechanisms can give rise to emergent collective behavior that cannot be inferred from isolated component performance alone.

To address this gap, we introduce Distributed Event-driven Scheduling Benchmark (DESBench), a benchmark for evaluating agent coordination in hierarchical, event-driven industrial scheduling, using it as a structured testbed for studying coordination in dynamic MAS. Our benchmark is built on a shared discrete-event simulation (DES) environment, where agents operate under partial observability, multi-timescale decision processes, and dynamically coupled constraints such as resource budgets and system capacity. Unlike prior environments \cite{zhu2025multiagentbench,wang2024battleagentbenchbench}, agent decisions directly influence the evolution of the underlying system, affecting future availability, resource usage, and constraint pressure. 

\begin{figure}
  \centering
  \includegraphics[width=0.9\textwidth]{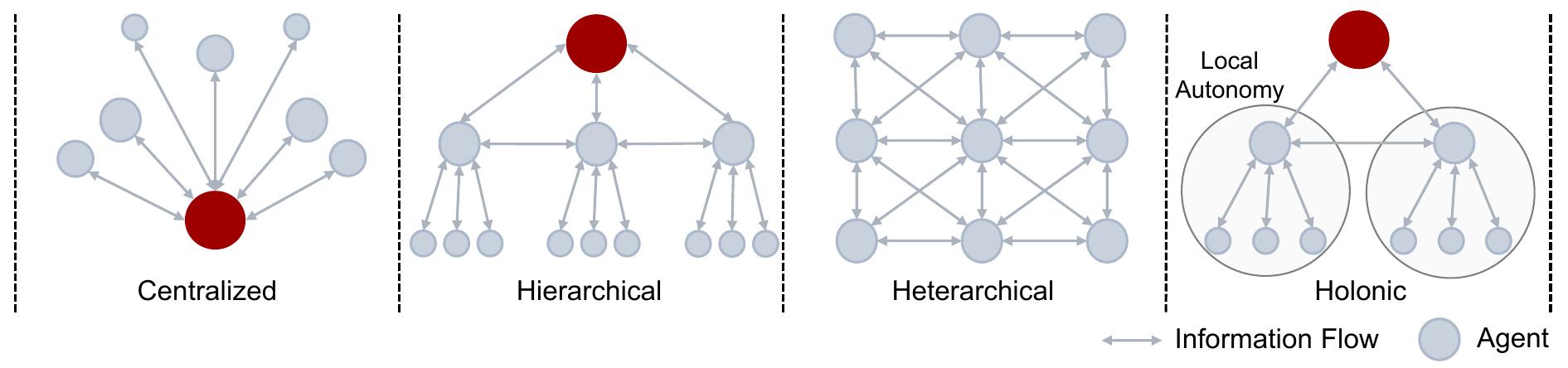}
  \caption{Four typical coordination paradigms: centralized, hierarchical, heterarchical, and holonic. Each represents a distinct mechanism of information flow, decision authority, and conflict resolution.}
  \label{intro}
\end{figure}

We focus on four representative coordination paradigms: centralized, hierarchical, heterarchical, and holonic, each corresponding to distinct mechanisms of information flow, decision authority, and conflict resolution (Figure \ref{intro}). These paradigms are instantiated as executable agent systems using orchestration frameworks, enabling systematic comparison within a unified environment. We design evaluation indices 
that measure multiple dimensions of agent performance, including effectiveness, constraint alignment, coordination efficiency, and robustness under disruptions. 

Our empirical results show that each paradigm exposes a distinct bottleneck: centralized coordination is robust but difficulty-sensitive, hierarchical coordination improves scheduling efficiency but risks level misalignment, heterarchical coordination gains adaptability at the price of heavy interaction, and holonic coordination preserves constraints but weakens global progress. These observations underscore that coordination design fundamentally governs system behavior in complex environments and point toward the need for adaptive and context‑aware coordination mechanisms in future research. Our work presents three key innovations:
\begin{enumerate}
    \item We introduce DESBench for evaluating agent coordination in event-driven industrial scheduling, capturing multi-timescale decision making and dynamically coupled constraints in shared and evolving environments.
    \item We implement four representative coordination paradigms, including centralized, hierarchical, heterarchical, and holonic, each characterized by distinct mechanisms of information flow, decision authority, and conflict resolution. We also provide a scaffold that enables the design and evaluation of new coordination strategies within our benchmark setting.
    \item We establish a unified evaluation framework with four metric dimensions, including effectiveness, constraint alignment, coordination efficiency, and robustness, and use it to reveal key trade-offs across coordination paradigms, providing insights for designing more adaptive and scalable coordination mechanisms in future MAS.
\end{enumerate}

\section{Related Work}

\textbf{Agent Coordination Mechanisms.} 
Agent coordination is fundamental in MAS, enabling distributed decision makers to share information and resolve conflicts toward collective goals \cite{sun2025multi}. Centralized coordination assigns a global controller to aggregate system-wide information and compute joint decisions. For example, EvoAgent \cite{dang2025multi} employs a centralized puppeteer to dynamically schedule and reprioritize agent actions during collaboration. While such designs can optimize global objectives under full observability, they scale poorly under uncertainty and partial observability \cite{li2025multi,xu2025centralized}. To address these limitations, recent work explores structured and decentralized coordination paradigms that balance global coherence with local autonomy \cite{ding2024multi}. Hierarchical coordination decomposes tasks into layered subproblems, enabling modular decision making and controlled cross-level interaction \cite{liu2025hm,tang2026hiva,zeng2025hierarchical}. For instance, SeqComm \cite{ding2024multi} combines inter-agent negotiation with top-down execution ordering across levels. Recent decentralized algorithms advance adaptive coordination through evolving interaction topologies and interaction-based reward design, as exemplified by AgentNet and CoMAS \cite{yang2025agentnet,xue2026comas}. Extensions incorporate adaptive clustering \cite{liu2025hygma} and privacy-aware knowledge sharing \cite{nalagatla2025hierarchical} to improve scalability while maintaining global consistency. Heterarchical coordination relies on peer-to-peer interaction and decentralized negotiation to reach consensus under local information constraints \cite{du2025contextual}, enhancing robustness in large-scale systems. Hybrid holonic coordination \cite{keramati2024hmlb} integrates hierarchical structure with decentralized autonomy, treating agents as self-contained units within cooperative collectives to enable dynamic reconfiguration and multi-level interaction.

\textbf{Benchmarking Coordination in MAS.} 
Existing MAS benchmarks predominantly emphasize performance on predefined scenarios and aggregate outcome measures, offering limited support for analyzing how coordination structures shape collective behavior. MultiAgentBench \cite{zhu2025multiagentbench} extends evaluation to collaboration and competition through milestone-based KPIs, while BattleAgentBench \cite{wang2024battleagentbenchbench} provides fine-grained assessment of cooperative and competitive capabilities. However, these benchmarks remain largely outcome-centric, focusing on task success or milestone achievement, and provide limited insight into coordination dynamics, communication efficiency, or sensitivity to environmental variation. Similarly, AgentsNet \cite{grotschla2025agentsnet} evaluates reasoning and topology use in MAS but does not explicitly isolate the impact of coordination mechanisms on system behavior. More generally, existing evaluation frameworks rely on final-output metrics, overlooking process-level signals such as interaction patterns or redundant collaboration paths \cite{lee2025gemmas}, despite evidence that similar outcome scores can mask substantial differences in internal coordination efficiency. These limitations align with recent surveys highlighting the lack of standardized protocols for evaluating dynamic coordination and emergent behavior in MAS \cite{shaikh2025llm,sun2025multi}. 

\section{Distributed Event-driven Scheduling Benchmark (DESBench)}
\label{sec:desbench}

\subsection{Problem Setting}
\begin{figure}
  \centering
  \includegraphics[width=\textwidth]{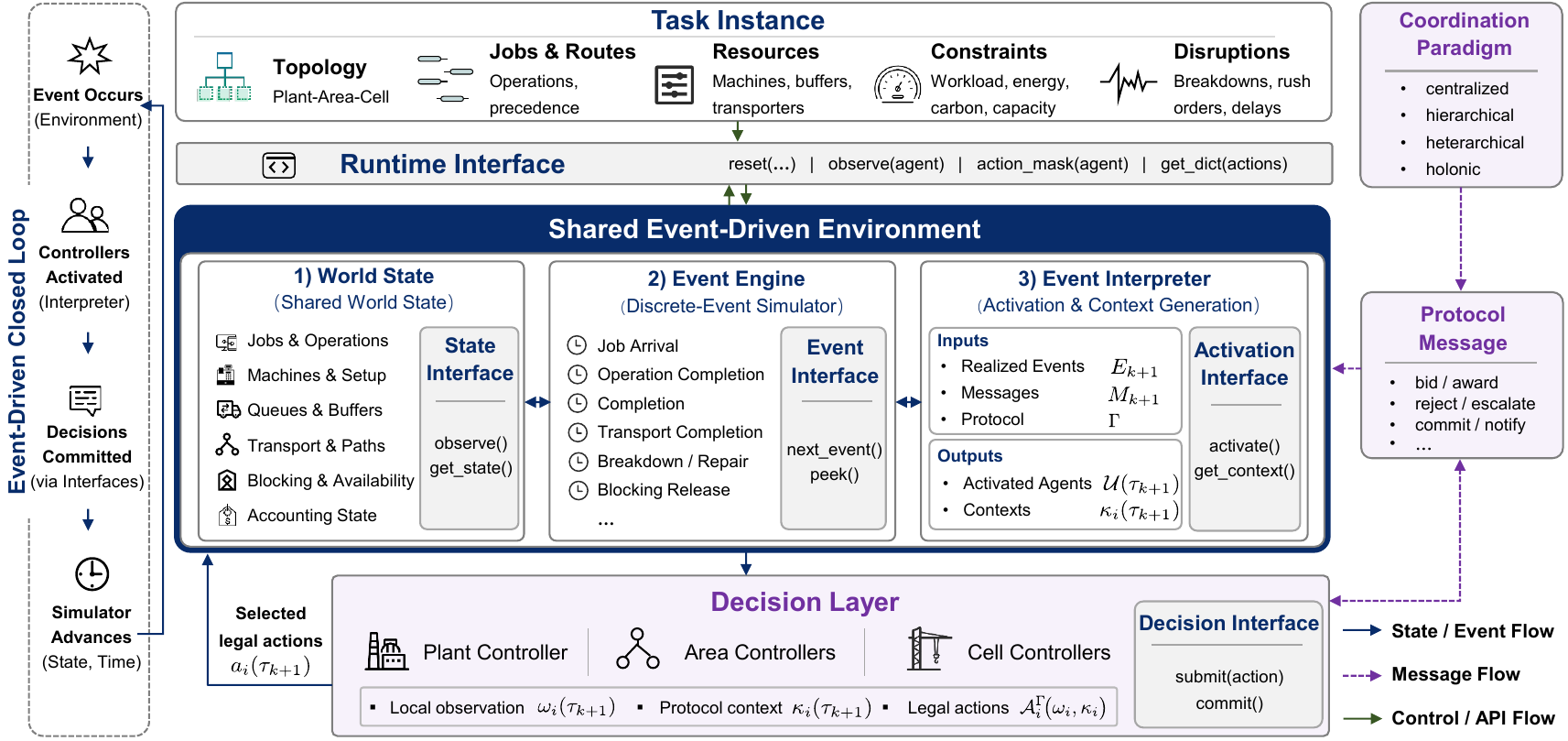}
    \caption{Overview of the DESBench. Agents at Plant, Area, and Cell levels receive local observations and protocol contexts, submit decisions via the interface, and advance the event engine. This unified architecture supports evaluation of multiple coordination paradigms under identical dynamics.}
  \label{main}
\end{figure}

DESBench studies industrial scheduling in a shared discrete-event environment, coordinating the execution of concurrent jobs. We consider a hierarchical MAS organized as a rooted coordination graph $\mathcal{H} = (\mathcal{V}, \mathcal{E}),$ where the vertex set $\mathcal{V}$ is partitioned into three decision levels, $\mathcal{V} = \mathcal{V}^{P} \cup \mathcal{V}^{A} \cup \mathcal{V}^{C},$ corresponding to the \emph{Plant}, \emph{Area}, and \emph{Cell} layers in a factory, respectively. A task instance specifies a fixed hierarchy $\mathcal{H}$ together with a shared discrete-event environment in which all agents interact through the evolving system state. Figure \ref{main} depicts the DESBench architecture, where a shared event-driven environment core drives state evolution and triggers agent activations under a specified coordination protocol. At each event epoch, activated agents observe local state and protocol context, submit decisions via runtime interfaces. 

At time $t$, the environment is in state $x_t \in \mathcal{X}$, which summarizes the current job set, machine status, queue contents, setup configuration, transport status, buffer occupancy, and resource-accounting variables. State transitions are not driven by synchronous time steps but by a stream of realized environment events. We define $E_k \subseteq \mathcal{E}_{\mathrm{env}}$ as the event set realized at time $\tau_k$, including job arrivals, operation completions, machine breakdowns, repairs, setup changes, transport completions, and blocking releases. These event sets induce a sequence of decision epochs $0 < \tau_1 < \tau_2 < \cdots < \tau_T,$ at which one or more agents are activated to make scheduling or coordination decisions. Between two consecutive decision epochs, the system evolves autonomously according to the underlying discrete-event dynamics.

A task contains a set of jobs $\mathcal{J}$, where each job $j \in \mathcal{J}$ is associated with an ordered processing route $r(j) = (o_{j1}, o_{j2}, \dots, o_{jm_j}),$ release conditions, and optionally due-date information. Each operation must be assigned to a feasible cell and then dispatched on a local machine subject to routing constraints, setup-dependent processing effects, transport delays across cells, finite inbound admission, and downstream blocking. Therefore, the scheduling process is jointly shaped by both local execution choices and higher-level coordination decisions.

At a decision epoch $\tau_k$, the realized event set $E_k$ is first incorporated into the event engine (see Section~\ref{subsec:shared_event_driven_environment}), yielding a pre-decision state $x_{\tau_k}^{-}$. One or more activated agents then issue coordination decisions, producing a joint decision $\mathbf{a}(\tau_k)$. Depending on the activated level, these decisions may correspond to backlog selection, area assignment, cell selection, local dispatch, rejection, rerouting, or escalation. After the activated agents commit their decisions, the event engine reaches a post-decision state $x_{\tau_k}^{+}$, from which the discrete-event dynamics advance to the next decision epoch. A joint execution over an episode thus yields an event-indexed trajectory
\begin{equation}
\xi =
\bigl(
x_0,
E_1, \tau_1, x_{\tau_1}^{-}, \mathbf{a}(\tau_1), x_{\tau_1}^{+},
\ldots,
E_T, \tau_T, x_{\tau_T}^{-}, \mathbf{a}(\tau_T), x_{\tau_T}^{+}
\bigr).
\end{equation}

A key feature of DESBench is that coordination is mediated through a shared system state rather than isolated subproblems. An action taken by one agent can change the future feasibility region of other agents by altering queue congestion, setup exposure, route availability, transport load, or budget pressure. Let $c(x_t) \in \mathbb{R}^d$ denote the vector of coupled operational quantities, including workload, energy use, carbon accumulation, and capacity utilization. Feasible control must respect resource and operational constraints $c(x_t) \preceq b,$ where $b \in \mathbb{R}^d$ is the task-specific budget. Violations need not occur immediately after an action, but may emerge later through delayed interactions in the event-driven dynamics, making coordination intrinsically long-horizon and cross-level. Formally, a DESBench instance is defined as
\begin{equation}
\mathcal{M} =
\bigl(
\mathcal{H},
\mathcal{X},
\mathcal{E}_{\mathrm{env}},
\{O_i\}_{i \in \mathcal{V}},
\{\mathcal{A}_i\}_{i \in \mathcal{V}},
P,
\mathcal{J},
b,
\Gamma
\bigr),
\end{equation}
where $\{O_i\}_{i \in \mathcal{V}}$ denotes the agent-specific observation operators over the shared state, $\{\mathcal{A}_i\}_{i \in \mathcal{V}}$ denotes the corresponding base action domains, $P$ denotes the event-driven transition law of the shared environment, and $\Gamma$ denotes the coordination protocol. In DESBench, different coordination paradigms correspond to different instantiations of $\Gamma$ over the same underlying task environment.


\subsection{Shared Event-Driven Environment}
\label{subsec:shared_event_driven_environment}


Given an instance $\mathcal{M}$, DESBench evolves as a shared asynchronous discrete-event system over the common state space $\mathcal{X}$. As shown in Figure~\ref{main}, we distinguish three runtime components: the shared world state $x_t$, a shared event engine $G$ that advances the simulator and realizes world events, and an event interpreter $I$ that maps newly realized world objects to agent activations under $\Gamma$. All agents interact through the same underlying world state $x_t$, and the environment advances only when discrete events are realized. In contrast to fixed-step control benchmarks \cite{8419722,wang2024moseac}, DESBench does not expose a universal synchronous decision loop. Instead, both the timing of decisions and the agents allowed to act are determined by the joint evolution of the physical environment and $\Gamma$.

At a decision epoch $\tau_k$, activated agents act on the pre-decision world state $x_{\tau_k}^{-}$ and produce a post-decision state $x_{\tau_k}^{+}$. The event engine then advances the shared simulator autonomously until the next decision-relevant stopping point. Let $E_{k+1} \subseteq \mathcal{E}_{\mathrm{env}}$ denote the set of environment events realized after decisions at epoch $\tau_k$ and incorporated at the next decision epoch $\tau_{k+1}$, and let $M_{k+1}$ denote the protocol-level objects generated during the same transition, including messages, commitments, and settlements. We write
\begin{equation}
G\bigl(x_{\tau_k}^{+}\bigr)
=
\bigl(x_{\tau_{k+1}}^{-}, E_{k+1}, \tau_{k+1}\bigr).
\label{engine}
\end{equation}
Thus, the event-engine $G$ is represented by the autonomous mapping that rolls the simulator forward and exposes the realized world-event set $E_{k+1}$. Over the same interval, the coordination runtime also produces protocol-level objects $M_{k+1}$, including messages, commitments, and settlements.

A decision epoch is not created at every state transition. Instead, the event interpreter $I$ consumes the newly realized $E_{k+1}$ and $M_{k+1}$ and determines both who acts next and what protocol context they receive. Let $\mathcal{U}(\tau_{k+1}) \subseteq \mathcal{V}$ denote the set of agents activated at epoch $\tau_{k+1}$. We write
\begin{equation}
I\bigl(E_{k+1}, M_{k+1}, x_{\tau_{k+1}}^{-}; \Gamma\bigr)=\Bigl(\mathcal{U}(\tau_{k+1}), \{\kappa_i(\tau_{k+1})\}_{i \in \mathcal{U}(\tau_{k+1})}\Bigr),
\label{interpre}
\end{equation}
where $\kappa_i(\cdot)$ denotes the protocol context. $I$ captures the key event-driven property of the benchmark: agents do not act continuously, but only when newly realized world events or protocol objects make a new decision legally necessary.

The resulting control process is inherently asynchronous and multi-level. Different agents may be activated for different reasons, including newly opened release opportunities, local execution events, budget alarms, breakdowns, repairs, parent-to-child allocations, or escalation messages. Consequently, DESBench couples physical evolution and coordination evolution within a unified runtime: \textit{environment events influence protocol activations, while protocol decisions in turn alter future environment events.}

At a decision epoch $\tau_k$, the realized event set $E_k$ has already been incorporated into the event engine, yielding the pre-decision state $x_{\tau_k}^{-}$. Each activated agent $i \in \mathcal{U}(\tau_k)$ receives a world-state-derived local observation $\omega_i(\tau_k) = O_i(x_{\tau_k}^{-})$, together with the protocol context $\kappa_i(\tau_k)$ returned by the event interpreter. The protocol context may encode the triggering event, a parent message, local commitment status, settlement outcome, or authority-conditioned decision semantics relevant to agent $i$ at $\tau_k$. Agent $i$ therefore selects an action from the protocol-conditioned feasible interface
\begin{equation}
a_i(\tau_k) \in \mathcal{A}_i^\Gamma\!\bigl(\omega_i(\tau_k), \kappa_i(\tau_k)\bigr). 
\end{equation}
Depending on the activated level and protocol instantiation, actions may correspond to backlog selection, area assignment, cell selection, local dispatch, rejection, rerouting, bid submission, or escalation. This representation makes explicit that DESBench does not expose a fixed, paradigm-independent decision menu: even under the same physical state, different instantiations of $\Gamma$ can induce different feasible actions because they alter who is activated, what information is revealed, and what authority or recovery options are available at that stage.

Finally, for each task instance, $\mathcal{X}$, $\mathcal{E}_{\mathrm{env}}$, $P$, $\mathcal{J}$, and $b$ are held constant, while only the coordination protocol $\Gamma$ varies. This isolates coordination as the sole experimental variable under a common dynamic substrate. Details of the shared event-driven runtime, activation logic, and state/accounting substrate are provided in Appendix~\ref{app:environment_runtime}. The event trajectories are driven by a set of real-world operational datasets \cite{TRENTESAUX20131204,DBLP:journals/corr/abs-1901-05669,hoss2025production}.

\subsection{Typical Coordination Paradigms}
\label{subsec:coordination_paradigms}
DESBench uses the coordination protocol $\Gamma$ as the primary experimental variable. For a fixed task instance, different coordination paradigms are represented as different protocol instantiations. We choose $\Gamma \in \left\{\Gamma^{\mathrm{cen}},\Gamma^{\mathrm{hie}}, \Gamma^{\mathrm{het}},\Gamma^{\mathrm{hol}}\right\}$ as typical examples. Each protocol $\Gamma$ specifies how authority is distributed across the Plant-Area-Cell hierarchy and how control flows after commitment, rejection, or failure. These dimensions determine the legal action sets $\mathcal{A}_i^\Gamma\!\bigl(\omega_i(\tau_k), \kappa_i(\tau_k)\bigr),$ the activation structure $\mathcal{U}(\tau_k)$, and the protocol-level objects $M_k$ generated along a trajectory. Thus, the same local observation can lead to different feasible decisions depending on the coordination paradigm.


\textbf{Centralized coordination.} Under $\Gamma^{\mathrm{cen}}$, public routing authority is concentrated at the Plant: after backlog selection, the Plant directly selects the execution Cell for the currently considered work item. Cell commitment is effectively accept-only, and rerouting is handled by Plant reselection rather than by Area-local recovery. Although the runtime still preserves the Plant-Area-Cell boundary internally, an Area commitment decision is not exposed as a separate public agent step in this mode.

\textbf{Hierarchical coordination.} Under $\Gamma^{\mathrm{hie}}$, decisions follow the fixed Plant-Area-Cell chain. The Plant selects an Area, the selected Area chooses a child Cell, and the Cell commits to execution under an accept-only contract interface. If a commitment cannot be settled, the failure propagates upward rather than being resolved through local sibling rerouting. This paradigm preserves a strict command hierarchy and tests whether structured decomposition improves coordination without granting strong local autonomy.

\textbf{Heterarchical coordination.} Under $\Gamma^{\mathrm{het}}$, allocation is performed through a mediated contract-net-style process. Candidate Cells may participate through a bid round and may reject contracts, while the runtime still preserves the Plant-Area-Cell object boundary. Thus, in the DESBench implementation, heterarchical coordination should be understood as a mediated heterarchical protocol rather than an unconstrained peer-to-peer system. It relaxes fixed top-down assignment by introducing competitive local participation, while retaining a bounded mediation structure for conflict resolution.

\textbf{Holonic coordination.} Under $\Gamma^{\mathrm{hol}}$, the system preserves the stable Plant-Area-Cell holarchy while granting bounded local autonomy. Cells may accept or reject contracts, and Areas may attempt local rerouting among eligible sibling Cells before escalating control upward. Plant-level fallback is used only after local recovery is exhausted. This paradigm combines hierarchical organization with local self-regulation, enabling DESBench to study whether bounded autonomy can improve robustness without losing the global coordination structure.

Table~\ref{tab:coordination_paradigms_main} summarizes the high-level distinction among the four paradigms. The exact implementation-level authority boundaries and prompts are provided in Appendix~\ref{app:coordination_protocol_details}.

\begin{table}[htbp]
\centering
\small
\caption{High-level coordination paradigms in DESBench. All paradigms share the same event-driven scheduling environment and differ only in the coordination protocol $\Gamma$.}
\label{tab:coordination_paradigms_main}
\begin{tabular}{p{0.10\linewidth} p{0.28\linewidth} p{0.25\linewidth} p{0.22\linewidth}}
\toprule
\textbf{Paradigm} & \textbf{Routing authority} & \textbf{Local autonomy} & \textbf{Conflict handling} \\
\midrule
Centralized
& Plant direct award  
& Cells execute without effective reject right
& Plant performs direct reselection \\
Hierarchical
& Fixed Plant-Area-Cell chain
& Cells are accept-only under commitment
& Reject or infeasibility escalates upward \\
Heterarchical
& Mediated bid/award process
& Cells can bid and reject within mediation
& Mediated settlement before fallback \\
Holonic
& Stable holarchy with bounded autonomy
& Cells can reject; Areas can locally reroute
& Area recovery before Plant fallback \\
\bottomrule
\end{tabular}
\end{table}

\subsection{Tasks and Evaluation Dimensions}
\label{subsec:tasks_and_evaluation_dimensions}

To systematically characterize performance beyond single objective scores, we organize the primary metrics into four evaluation dimensions: \emph{effectiveness}, \emph{constraint alignment}, \emph{coordination efficiency}, and \emph{robustness}. Effectiveness measures schedule quality over the event trajectory; constraint alignment captures how well runs respect shared operational limits; coordination efficiency quantifies interaction cost and decision overhead induced by a protocol; and robustness distinguishes among qualitatively different failure modes under unresolved workloads or episode truncation. Table~\ref{tab:main_metric_dimensions} summarizes the main metrics, with the full benchmark catalog provided in Appendix~\ref{app:metric_catalog}.

\begin{table*}[htbp]
\centering
\small
\caption{Main evaluation metrics in DESBench and their interpretations.}
\label{tab:main_metric_dimensions}
\begin{tabular}{
  c                
  p{0.06\linewidth} 
  p{0.72\linewidth}}
\toprule
\textbf{Dimension} & \textbf{Metric} & \textbf{Interpretation} \\
\midrule

\multirow{3}{1.8cm}{Effectiveness \\ ($\mathcal{G}_{\mathrm{eff}}$)}
& $G_{\mathrm{mk}}$ & \texttt{Mean makespan}: average final simulation time at episode termination. \\
& $G_{\mathrm{td}}$ & \texttt{Tardiness}: episode-level total tardiness over completed jobs with due dates. \\
& $G_{\mathrm{cj}}$ & \texttt{Completed jobs}: average number of jobs completed within the episode. \\
\midrule
\multirow{4}{1.8cm}{Constraint Alignment \\ ($\mathcal{G}_{\mathrm{con}}$)} 
& $G_{\mathrm{vc}}$ & \texttt{Violation count}: recorded overshoot count from violation summaries. \\
& $G_{\mathrm{eo}}$ & \texttt{Energy overshoot magnitude}: average cumulative amount by which energy usage exceeds the budget. \\
& $G_{\mathrm{ed}}$ & \texttt{Energy overshoot duration}: cumulative time spent in energy overshoot. \\
& $G_{\mathrm{en}}$ & \texttt{Energy}: mean total energy usage per episode. \\
\midrule

\multirow{4}{1.8cm}{Coordination Efficiency \\ ($\mathcal{G}_{\mathrm{coord}}$)} 
& $G_{\mathrm{ds}}$ & \texttt{Decision steps}: mean number of executed rollout decision steps per episode. \\
& $G_{\mathrm{cm}}$ & \texttt{Coordination messages}: mean number of protocol-level coordination messages per episode. \\
& $G_{\mathrm{fb}}$ & \texttt{Fallback‑contract decisions}: mean number of framework decisions attributed to the fallback-contract source. \\
& $G_{\mathrm{llm}}$ & \texttt{LLM-driven decision count}: mean number of framework decisions attributed to the LLM source. \\
\midrule

\multirow{4}{1.8cm}{Robustness \\ ($\mathcal{G}_{\mathrm{rob}}$)} 
& $G_{\mathrm{tr}}$ & \texttt{Truncation rate}: fraction of episodes marked as truncated. \\
& $G_{\mathrm{dj}}$ & \texttt{Completion debt (jobs)}: average number of unfinished jobs at episode end. \\
& $G_{\mathrm{ut}}$ & \texttt{Unfinished overdue tardiness}: mean overdue tardiness debt from unfinished jobs at episode end. \\
& $G_{\mathrm{do}}$ & \texttt{Completion debt (operations)}: average number of remaining unprocessed operations at episode end. \\

\bottomrule
\end{tabular}
\end{table*}

\textbf{Effectiveness.} 
$\mathcal{G}_{\mathrm{eff}}$ quantifies schedule quality. Specifically, $G_{\mathrm{mk}}$ records final simulation time at episode termination, $G_{\mathrm{td}}$ measures episode-level total tardiness over completed jobs with due dates, and $G_{\mathrm{cj}}$ counts completed jobs. 

\textbf{Constraint Alignment.}  
Metrics in $\mathcal{G}_{\mathrm{con}}$ assess how consistently a coordination paradigm operates within shared operational limits induced by the environment. $G_{\mathrm{vc}}$ is derived from runtime violation summaries, with an excess-state fallback when structured summaries are unavailable. The remaining metrics quantify energy-specific breach severity and resource use, with $G_{\mathrm{eo}}$ measuring cumulative energy overshoot magnitude, $G_{\mathrm{ed}}$ measuring cumulative time spent in energy overshoot, and $G_{\mathrm{en}}$ reporting mean total energy usage per episode.

\textbf{Coordination Efficiency.}
$\mathcal{G}_{\mathrm{coord}}$ targets the cost of coordination itself under a given protocol $\Gamma$, complementing outcome quality with indicators of control activity, communication burden, and audited decision provenance. $G_{\mathrm{ds}}$ counts executed rollout decision steps, and $G_{\mathrm{cm}}$ counts protocol-level coordination messages exposed by the runtime. $G_{\mathrm{fb}}$ and $G_{\mathrm{llm}}$ are framework-audited decision-source counts, recording how often a decision is attributed to fallback and LLM, respectively.

\textbf{Robustness.}
$\mathcal{G}_{\mathrm{rob}}$ characterizes early termination and residual work debt at episode end. Truncation rate ($G_{\mathrm{tr}}$) records episodes marked as truncated, which may arise from deadlock termination or no-progress stopping. The remaining metrics quantify unresolved workload through unfinished jobs ($G_{\mathrm{dj}}$), overdue debt from unfinished jobs ($G_{\mathrm{ut}}$), and remaining operations ($G_{\mathrm{do}}$).

\section{Experimental Result and Discussion}
\label{exp}
%
\textbf{Experimental Setting.}
We evaluate four coordination paradigms, each implemented in two orchestration frameworks, LangGraph \cite{wang2024agent} and AgentScope \cite{gao2024agentscope}, and paired with three representative LLMs: GPT-5.4, Gemini-3-Flash, and Qwen-3.5. The experiments adopt A5C12 topology, consisting of five Areas and twelve Cells, partitioned as 4-3-2-2-1. Three scenario profiles are tested, each designed to emphasize a different coordination difficulty: branch pressure (high task branching and routing decisions), strong cluster pull (tight coupling of resource usage across areas), and late task commitments (delayed task scheduling with potential resource conflicts). Each paradigm is evaluated with a total of 30 test episodes, with 10 test seeds per framework and LLM pair for each of the three scenarios. The final results for each paradigm are averaged across the three scenario profiles. Experiments are conducted on a local Windows workstation equipped with an Intel Core i7-12700K CPU and 32~GB RAM. The evaluation does not require specialized hardware and has modest computational requirements since it focuses on coordination behavior rather than large‑scale model training. Implementation details are provided in Appendix~\ref{ia}.

\textbf{Experimental Result.} We begin by examining the numerical results across different LLMs and framework implementations. As shown in Table~\ref{tab:main_benchmark_matrix}, coordination structure remains the dominant source of variation, while LLM and framework differences are usually secondary. The table shows that switching coordination paradigms changes system behavior more substantially.


\newcommand{\blankmetriccells}{& & & & & & & & & & & & & & &}
\begin{table*}[htbp]
\centering
\setlength{\tabcolsep}{3pt}
\renewcommand{\arraystretch}{1.08}
\caption{Main benchmark comparison matrix across coordination paradigms.}
\label{tab:main_benchmark_matrix}
\resizebox{\textwidth}{!}{%
\begin{tabular}{lll*{15}{c}}
\toprule
\multirow{3}{*}{\textbf{Paradigm}} & \multirow{3}{*}{\textbf{Framework}} & \multirow{3}{*}{\textbf{LLM}} & \multicolumn{15}{c}{\textbf{Evaluation Metrics}} \\
\cmidrule(lr){4-18}
& & & \multicolumn{3}{c}{$\mathcal{G}_{\mathrm{eff}}$} & \multicolumn{4}{c}{$\mathcal{G}_{\mathrm{con}}$} & \multicolumn{4}{c}{$\mathcal{G}_{\mathrm{coord}}$} & \multicolumn{4}{c}{$\mathcal{G}_{\mathrm{rob}}$} \\
\cmidrule(lr){4-6}\cmidrule(lr){7-10}\cmidrule(lr){11-14}\cmidrule(lr){15-18}
& & & $G_{\mathrm{mk}} \downarrow$ & $G_{\mathrm{td}}\downarrow$ & $G_{\mathrm{cj}}\uparrow$ & $G_{\mathrm{vc}}\downarrow$ & $G_{\mathrm{eo}}\downarrow$ & $G_{\mathrm{ed}}\downarrow$ & $G_{\mathrm{en}}\downarrow$ & $G_{\mathrm{ds}}\downarrow$ & $G_{\mathrm{cm}}\downarrow$ & $G_{\mathrm{fb}}\downarrow$ & $G_{\mathrm{llm}}\downarrow$ & $G_{\mathrm{tr}}\downarrow$ & $G_{\mathrm{dj}}\downarrow$ & $G_{\mathrm{ut}}\downarrow$ & $G_{\mathrm{do}}\downarrow$ \\
\midrule

\multirow{6}{*}{$\Gamma^{\mathrm{cen}}$}
& \multirow{3}{*}{LangGraph} & GPT-5.4 & 78.2 & 255.6 & 9.8 & 32.4 & 14471.5 & 9.7 & 12245.0 & 118.0 & 118.0 & 0.1 & 117.9 & 0.03 & 0.2 & 1.7 & 0.8 \\
& & Gemini-3-Flash & 75.4 & 247.9 & 10.0 & 32.7 & 11947.0 & 7.0 & 11828.9 & 120.0 & 120.0 & 1.1 & 118.9 & 0.00 & 0.0 & 0.0 & 0.0 \\
& & Qwen-3.5 & 74.3 & 247.7 & 9.8 & 31.4 & 12129.5 & 7.1 & 11675.8 & 118.3 & 118.3 & 0.0 & 118.3 & 0.03 & 0.2 & 2.0 & 0.7 \\
\cmidrule(lr){2-18}
& \multirow{3}{*}{AgentScope} & GPT-5.4 & 76.3 & 252.1 & 9.4 & 31.8 & 14128.4 & 8.9 & 11943.2 & 115.4 & 115.3 & 0.1 & 115.3 & 0.07 & 0.6 & 3.4 & 1.8 \\
& & Gemini-3-Flash & 75.4 & 247.9 & 10.0 & 32.7 & 11946.9 & 7.0 & 11828.8 & 120.0 & 120.0 & 1.1 & 118.9 & 0.00 & 0.0 & 0.0 & 0.0 \\
& & Qwen-3.5 & 76.2 & 251.0 & 9.8 & 31.6 & 12869.8 & 8.9 & 11953.4 & 118.3 & 118.3 & 0.0 & 118.3 & 0.03 & 0.2 & 2.0 & 0.7 \\
\midrule

\multirow{6}{*}{$\Gamma^{\mathrm{hie}}$}
& \multirow{3}{*}{LangGraph} & GPT-5.4 & 66.5 & 156.9 & 9.2 & 57.2 & 15292.6 & 36.0 & 10669.2 & 190.7 & 115.1 & 0.2 & 191.5 & 0.13 & 0.8 & 7.9 & 2.1 \\
& & Gemini-3-Flash & 62.0 & 143.9 & 9.7 & 54.7 & 12461.3 & 29.9 & 10018.2 & 196.4 & 118.4 & 0.5 & 196.8 & 0.07 & 0.3 & 4.0 & 0.8 \\
& & Qwen-3.5 & 58.2 & 122.9 & 7.7 & 44.9 & 12318.6 & 28.6 & 9344.7 & 172.5 & 104.5 & 0.0 & 174.0 & 0.37 & 2.3 & 15.1 & 6.5 \\
\cmidrule(lr){2-18}
& \multirow{3}{*}{AgentScope} & GPT-5.4 & 64.3 & 151.8 & 8.9 & 55.2 & 13644.9 & 33.8 & 10317.0 & 186.4 & 112.5 & 0.1 & 187.4 & 0.20 & 1.1 & 12.3 & 3.2 \\
& & Gemini-3-Flash & 62.0 & 143.9 & 9.7 & 54.8 & 12461.2 & 29.8 & 10012.2 & 196.4 & 118.4 & 0.6 & 196.8 & 0.07 & 0.3 & 4.0 & 0.8 \\
& & Qwen-3.5 & 57.8 & 124.5 & 7.9 & 46.6 & 12596.8 & 28.7 & 9285.6 & 172.7 & 104.6 & 0.0 & 174.2 & 0.33 & 2.1 & 10.3 & 6.4 \\
\midrule

\multirow{6}{*}{$\Gamma^{\mathrm{het}}$}
& \multirow{3}{*}{LangGraph} & GPT-5.4 & 64.4 & 154.0 & 10.0 & 57.7 & 10540.8 & 31.1 & 10329.6 & 331.1 & 471.2 & 1.8 & 330.3 & 0.00 & 0.0 & 0.0 & 0.0 \\
& & Gemini-3-Flash & 62.5 & 148.8 & 9.8 & 55.7 & 12653.7 & 30.2 & 10099.6 & 201.3 & 244.9 & 0.3 & 202.0 & 0.03 & 0.2 & 1.4 & 0.4 \\
& & Qwen-3.5 & 62.7 & 138.3 & 8.6 & 51.8 & 13341.8 & 31.9 & 10070.0 & 195.6 & 243.0 & 0.0 & 197.1 & 0.27 & 1.4 & 13.8 & 3.7 \\
\cmidrule(lr){2-18}
& \multirow{3}{*}{AgentScope} & GPT-5.4 & 61.7 & 145.5 & 9.6 & 50.9 & 8974.6 & 27.8 & 9912.2 & 324.0 & 461.3 & 1.8 & 323.2 & 0.10 & 0.4 & 6.1 & 1.2 \\
& & Gemini-3-Flash & 62.3 & 148.9 & 9.8 & 56.1 & 12695.0 & 30.1 & 10077.0 & 201.5 & 245.1 & 0.3 & 202.2 & 0.03 & 0.2 & 1.4 & 0.4 \\
& & Qwen-3.5 & 69.0 & 150.5 & 9.3 & 58.0 & 16163.4 & 38.4 & 11026.8 & 200.3 & 247.4 & 0.0 & 201.8 & 0.13 & 0.7 & 5.7 & 2.0 \\
\midrule

\multirow{6}{*}{$\Gamma^{\mathrm{hol}}$}
& \multirow{3}{*}{LangGraph} & GPT-5.4 & 25.9 & 30.7 & 2.8 & 5.3 & 738.8 & 2.8 & 4200.3 & 128.8 & 90.3 & 0.0 & 129.8 & 0.93 & 7.2 & 17.7 & 23.3 \\
& & Gemini-3-Flash & 58.3 & 125.3 & 9.0 & 51.0 & 10660.1 & 26.4 & 9434.1 & 194.3 & 118.8 & 0.6 & 194.7 & 0.27 & 1.0 & 13.7 & 2.4 \\
& & Qwen-3.5 & 27.4 & 38.9 & 3.1 & 12.9 & 3092.7 & 7.5 & 4417.0 & 101.0 & 63.8 & 0.0 & 102.3 & 0.83 & 6.9 & 8.9 & 23.4 \\
\cmidrule(lr){2-18}
& \multirow{3}{*}{AgentScope} & GPT-5.4 & 28.7 & 38.1 & 3.4 & 5.2 & 407.9 & 1.8 & 4672.5 & 141.6 & 97.9 & 0.0 & 142.7 & 1.00 & 6.6 & 30.4 & 20.8 \\
& & Gemini-3-Flash & 58.3 & 125.3 & 9.0 & 51.0 & 10660.1 & 26.4 & 9434.1 & 194.3 & 118.8 & 0.6 & 194.7 & 0.27 & 1.0 & 13.7 & 2.4 \\
& & Qwen-3.5 & 30.9 & 47.4 & 3.8 & 16.2 & 4400.8 & 9.8 & 4991.1 & 110.8 & 70.0 & 0.0 & 112.2 & 0.77 & 6.2 & 6.0 & 21.3 \\

\bottomrule
\end{tabular}%
}
\end{table*}

Centralized coordination excels in simple, well-structured environments. Due to central decision-making, it maintains strong robustness in this benchmark setting, as reflected by its favorable performance in the robustness metrics (\(\mathcal{G}_{\mathrm{rob}}\)). Additionally, its message passing is relatively sparse, as evidenced by the lower values of \( G_{\mathrm{ds}} \) and \( G_{\mathrm{cm}} \). However, its task processing efficiency is relatively low, as indicated by higher values of \( G_{\mathrm{mk}} \) and \( G_{\mathrm{td}} \). Moreover, as the complexity of the environment increases, the performance of centralized coordination degrades significantly (see Appendix~\ref{ae}).

Hierarchical coordination simplifies problems by decomposing tasks and assigning them to different levels of hierarchy. Its hierarchical strategy ensures stable robustness (a solid performance in \(\mathcal{G}_{\mathrm{rob}}\)), though higher violations are observed in certain scenarios due to potential delays or misalignments between different hierarchical levels, which is a key issue of this paradigm. The improved task decomposition significantly enhances efficiency (\( \mathcal{G}_{\mathrm{eff}} \)) compared to centralized coordination. Coordination costs (\( \mathcal{G}_{\mathrm{coord}} \)) in hierarchical coordination are higher than in centralized coordination but lower than in heterarchical coordination. This makes hierarchical coordination a balanced approach for environments of moderate complexity.

Heterarchical coordination offers greater flexibility and adaptability, dynamically adjusting to different task requirements. However, its violation rate (\( \mathcal{G}_{\mathrm{con}} \)) is not low, which is due to the increased complexity and the need for constant adjustments between levels. Efficiency (\( \mathcal{G}_{\mathrm{eff}} \)) reaches a moderate level, and robustness is strong because the system can respond to changes and failures more effectively through its flexible structure. However, it incurs the highest coordination and message counts (\( \mathcal{G}_{\mathrm{coord}} \)). For instance, the decision steps ($G_{\mathrm{ds}}$) are 331.1, and the coordination message count ($G_{\mathrm{cm}}$) is 471.2, reflecting an increased communication overhead compared to both centralized and hierarchical models. While heterarchical coordination provides flexibility and robustness, its high coordination cost reduces its overall efficiency. Excessive communication can lead to system confusion, hindering the effectiveness of the coordination process.

Holonic coordination emphasizes autonomy by allowing individual components to make local decisions. It achieves low delay-related effectiveness. The low makespan ($G_{\mathrm{mk}}$) and tardiness ($G_{\mathrm{td}}$) partly result from limited completed workload ($G_{\mathrm{cj}}$) rather than uniformly superior scheduling quality. Holonic coordination attains the best constraint satisfaction. This is because autonomous local control tends to act conservatively and avoids aggressive global resource contention. Its coordination efficiency is also competitive, with moderate decision steps ($G_{\mathrm{ds}}$) and coordination messages ($G_{\mathrm{cm}}$), which are far lower than those of heterarchical coordination. However, holonic coordination shows the weakest robustness, as reflected by a high truncation rate ($G_{\mathrm{tr}}$), large job completion debt ($G_{\mathrm{dj}}$), and high remaining operation debt ($G_{\mathrm{do}}$). This weakness arises from the fragmentation introduced by autonomy: local units can avoid violations, but they may fail to maintain global progress and recovery under dynamic disruptions, making the paradigm less effective in dynamic environments. It is worth noting that performance varies across different LLMs, as this paradigm relies on local autonomy, which amplifies model-specific behaviors.


Additional experiments on instances, scalability, and task difficulty are provided in Appendix~\ref{ae}.


\section{Discussion and Future Directions}
The above results expose a fundamental and non-trivial trade-off landscape. Only four typical static coordination paradigms already reveal substantial behavioral diversity. No coordination paradigm dominates across all dimensions and each paradigm occupies a distinct region in the space. This highlights that coordination is not merely an implementation choice, but a first-order design variable that directly shapes system behavior.

A key insight is the inherent tension between \emph{global consistency} and \emph{local responsiveness}. Centralized coordination maintains strong robustness and low coordination overhead, yet suffers from scalability and efficiency limitations as system complexity grows. In contrast, heterarchical coordination achieves adaptability and resilience at the cost of excessive communication, revealing that unstructured flexibility can introduce coordination noise that degrades system-level efficiency. Hierarchical coordination partially mitigates this tension by introducing structure, but its reliance on cross-level alignment creates systematic vulnerabilities. Holonic coordination, while excelling in constraint satisfaction, exposes a critical failure mode: local autonomy can fragment global progress, leading to poor robustness despite favorable local metrics.

In addition, these results also show that single-metric comparisons are inadequate for evaluating coordination. Metrics such as makespan or tardiness can be misleading when completion and truncation are ignored, as illustrated by the holonic results in Table~\ref{tab:main_benchmark_matrix}. Looking forward, several research directions emerge. First, hybrid coordination mechanisms \cite{sun2025multi,moore2025taxonomy} that combine the strengths of multiple paradigms (e.g., hierarchical structures with adaptive heterarchical interactions) represent a promising avenue for balancing efficiency and robustness. Second, learning-based approaches that adapt coordination strategies online based on system state and workload characteristics could mitigate the rigidity observed in fixed protocols \cite{dang2025multi,yue2025sequential}. Third, improving communication efficiency \cite{kwon2025cp,leong2025amas}, for example through selective activation or information compression, is essential to address the scalability bottlenecks observed in heterarchical systems. 

Advancing MAS toward real-world complex tasks such as industrial scheduling requires rethinking coordination not as a static design, but as a dynamic and context-dependent process. A systematic understanding of these principles can help guide the development of next-generation coordination frameworks and stimulate deeper investigation into the interplay between control, communication, and system dynamics. Finally, a guidance on how to use our benchmark is provided in Appendix~\ref{ues}.

\section{Conclusion}
This paper introduces DESBench, a unified benchmark for evaluating agent coordination in distributed event-driven scheduling. Our results reveal that coordination design fundamentally shapes system behavior, with clear trade-offs across effectiveness, constraint alignment, coordination efficiency, and robustness. No single paradigm dominates and each exposes distinct strengths and limitations under shared dynamics. We expect DESBench to provide a principled foundation for studying agent coordination mechanisms and to facilitate the development of more adaptive, efficient, and robust coordination strategies in multi-agent systems.




\bibliography{ref} 
\bibliographystyle{unsrtnat}

\appendix





\section{Additional Environment and Runtime Details}
\label{app:environment_runtime}

One decision epoch unfolds in four stages:
\begin{enumerate}
    \item Starting from the post-decision state $x_{\tau_k}^{+}$, the event engine advances autonomously until the next decision-relevant stopping point is reached, yielding a new pre-decision state $x_{\tau_{k+1}}^{-}$ together with realized world events $E_{k+1}$.
    \item The event interpreter combines these world events with any newly materialized protocol-level objects $M_{k+1}$ and computes the next activated set $\mathcal{U}(\tau_{k+1})$.
    \item For each activated agent $i \in \mathcal{U}(\tau_{k+1})$, the runtime constructs the benchmark-visible decision boundary (local observation), a protocol context, and a protocol-conditioned legal action set.
    \item Returned actions are validated, committed to the event engine, and written back as updated state and protocol artifacts. 
\end{enumerate}


\subsection{Event Engine}
\label{app:event_engine}

The event engine is the shared discrete-event scheduling substrate used by all coordination protocols. It advances physical production time, realizes environment events, updates job, machine, buffer, transport, blocking, and accounting states, and exposes the next decision-relevant stopping point. All coordination paradigms share this same substrate: the event queue, job/machine/buffer/transport/blocking semantics, setup and failure processes, and resource-accounting logic are held fixed. Different paradigms alter only the protocol layer $\Gamma$, which changes agent activation, information exposure, authority boundaries, and legal actions. The substrate follows the production-scheduling and discrete-event simulation in which job release, operation completion, setup, transport, buffer admission, blocking, and disruptions are represented as event-driven state changes \cite{TRENTESAUX20131204}. 

DESBench is therefore not exposed to controllers (agents) as a fully observed fixed-step Markov decision process. The full internal simulator state includes the pending event queue, machine status, job progress, ready jobs, blocked backlog, buffer occupancy, transport timers, residual processing and setup times, protocol artifacts, random generator state, and accumulated accounting variables. This complete state is internal to the engine. At benchmark level, agents are activated only at irregular event epochs and receive local observations, protocol contexts, and protocol-conditioned legal action sets.

In this case, Equation \ref{engine} is implemented by the event engine's autonomous advancement routine. $G$ is the shared rollout operator that advances the physical DES from the validated post-decision state $x_{\tau_k}^{+}$ to the next exposed pre-decision state while accumulating the realized environment-event set $E_{k+1}$ and the corresponding timestamp $\tau_{k+1}$.

\paragraph{Autonomous event-advance loop.}
Starting from $x_{\tau_k}^{+}$, the engine (i) selects the earliest pending environment event from the deterministic queue; (ii) advances the simulator clock to that event time; (iii) applies the event-specific physical transition, which can be written as $x \leftarrow \Phi_e(x)$; (iv) updates the affected machine, job, buffer, transport, blocking, and accounting variables and schedules any induced follow-on events; (v) appends the realized event via $E_{k+1} \leftarrow E_{k+1} \cup \{e\}$; and (vi) continues until a decision-relevant stopping point is reached. The stopping point is not simply the next queue pop, but the first runtime boundary at which a new agent intervention is legally required or the episode terminates. Thus, a physical event is not identical to a decision epoch: several world events may be processed before any agent is activated.

The engine maintains a deterministic event queue. Events are primarily ordered by event time, with deterministic tie-breaking for events realized at the same timestamp. The event alphabet shared across paradigms includes both basic production events and inter-cell transport/buffering events. The base simulator handles the following events:
\begin{itemize}
    \item \texttt{arrival}, \texttt{setup\_start}, \texttt{setup\_complete}, \texttt{finish}, \texttt{breakdown}, and \texttt{repair}.
\end{itemize}
The inter-cell scheduling layer extends this alphabet with the following events 
\begin{itemize}
    \item \texttt{shared\_backlog\_arrival}, \texttt{transport\_start}, \texttt{transport\_complete}, \texttt{stage\_release\_ready}, \texttt{buffer\_wait\_start}, \texttt{blocking\_start}, \texttt{buffer\_admit}, and \texttt{blocking\_end}.
\end{itemize}
These labels describe realized world events in the shared substrate. They are distinct from protocol-level objects such as messages, commitments, settlements, rejection records, or escalation artifacts generated by the coordination layer. These event classes mirror common real-world scheduling constraints such as transport logistics, buffer management, sequence/setup effects, machine breakdowns, and stochastic processing conditions \cite{hoss2025production}.

The main scheduling transitions follow standard discrete-event shop-floor semantics. A job-arrival event inserts work into the ready set when it is eligible and not already processing or completed. A dispatch decision binds a ready job to a feasible idle machine. Once committed, dispatch proceeds as
\[
\texttt{ready job}
\;{\rightarrow\;}
\texttt{feasible idle machine}
\;{\rightarrow\;}
\texttt{setup if needed}
\;{\rightarrow\;}
\texttt{processing}
\;{\rightarrow\;}
\texttt{finish}.
\]
If the selected machine requires a setup change, the engine first realizes setup-related events before processing begins; otherwise, processing starts immediately and a finish event is scheduled. A \texttt{finish} event frees the machine, advances the job's operation index, and then either marks the job as completed or releases it to the next route stage. Completion therefore updates both physical availability and accounting state, while nonterminal completion creates the next routing, transport, or stage-release opportunity.

The inter-cell layer introduces transport, finite inbound admission, and downstream blocking. A routing or assignment decision does not immediately make a job available for processing. Instead, the job enters a transport process governed by transport-related events. Only upon transport completion does it enter the destination cell's inbound state and become eligible for local setup and dispatch. Similarly, a \texttt{stage\_release\_ready} event indicates that a downstream release has become physically admissible, but not yet processed. When a downstream inbound buffer is full, the engine neither drops the job nor treats the assignment as completed. Instead, it records buffer waiting or blocking and moves the item into a blocked backlog. Once downstream capacity becomes available, admission or blocking-release events are triggered, allowing the job to re-enter the appropriate backlog or inbound state and resume normal routing and dispatch.

Failure and repair are also treated as world events. A \texttt{breakdown} changes the availability of the affected machine and prevents the corresponding processing slice from completing normally under the simulator's failure semantics. The machine remains unavailable until the matching \texttt{repair} event is realized. After repair, the engine either resumes eligible interrupted work or returns the machine to the idle state. This is why routing decisions in DESBench do not immediately convert into processing availability: transport, finite inbound capacity, and failure recovery are modeled as separate physical events. During these transitions, the engine records the accounting and trace substrate used by the benchmark metrics. It accumulates simulation time, energy usage, carbon emissions, setup time, transport time, blocking time, buffer waiting time, completed-job counts, remaining operations, and episode-status indicators such as truncation, deadlock, and done flags. These traces make the benchmark auditable: the reported metrics are derived from the same shared event trajectory.

\subsection{Event Interpreter}
\label{app:event_interpretation}
Once the event engine has produced the next pre-decision state, the runtime switches from physical advancement to activation logic. The interpreter takes as input the realized world-event set $E_{k+1}$, the newly materialized protocol-level object set $M_{k+1}$, the pre-decision state $x_{\tau_{k+1}}^{-}$, and the current coordination protocol $\Gamma$, and maps them to the next activated set and controller-facing protocol contexts. As described in Equation \ref{interpre} in Section~\ref{subsec:shared_event_driven_environment}, DESBench distinguishes between \emph{environment events} and \emph{protocol-level coordination objects}. The former are generated by the shared event engine and capture changes in the underlying operational system, while the latter arise from the execution of the coordination protocol itself. We write
\begin{equation}
E_{k+1} \subseteq \mathcal{E}_{\mathrm{env}},
\qquad
M_{k+1} \subseteq \mathcal{M}_{\mathrm{proto}},
\end{equation}
where $\mathcal{M}_{\mathrm{proto}}$ denotes the space of materialized protocol-level objects, including messages, commitments, allocations, settlements, rejection records, and escalation artifacts. 

Operationally, $I$ scans the realized event set $E_{k+1}$ and protocol objects $M_{k+1}$ for decision triggers, and resolves them under the authority, rerouting, escalation, and commitment semantics of $\Gamma$. The output is not merely a set of agent identifiers: $\mathcal{U}(\tau_{k+1})$ specifies which agents are obligated to act, while each $\kappa_i(\tau_{k+1})$ provides a role-conditioned context constructed from triggering events, relevant protocol objects, and necessary local state summaries. The interpreter maps these world events and protocol-level objects to activation records. We denote the activation record for agent $i$ at epoch $\tau_k$ by $\alpha_i(\tau_k)$ and write it as
\begin{equation}
\alpha_i(\tau_k)
=
\bigl(
 i,\;
 \mathrm{cause}_i(\tau_k),\;
 \mathrm{loop}_i(\tau_k),\;
 \mathrm{time}_i(\tau_k),\;
 \mathrm{ctx}_i(\tau_k)
\bigr).
\end{equation}
where $\mathrm{cause}_i(\tau_k)$ records whether the activation is induced by world events or protocol-level objects, $\mathrm{loop}_i(\tau_k)$ identifies the current functional loop of the runtime, and $\mathrm{ctx}_i(\tau_k)$ stores the structured context attached to that activation. This activation record is an implementation-level representation used to build $\kappa_i(\tau_k)$. In the current runtime, activations arise primarily from two sources:
\begin{enumerate}
    \item \textbf{World-event activations}, induced by newly realized environment events such as release opportunities, completion-triggered routing, queue-state changes, budget alarms, breakdowns, or repairs.
    \item \textbf{Protocol-object activations}, induced by coordination objects addressed to a specific agent, such as assignment requests, parent-to-child allocations, child feedback, settlement notices, rejection records, or escalation requests.
\end{enumerate}
Only agents in $\mathcal{U}(\tau_k)$ are required to act at epoch $\tau_k$, and the activated set itself is an output of the event interpreter.

\phantomsection\label{app:controller_interface}
\paragraph{Controller-facing payload.}
Conditional on activation, agent $i$ does not observe the full hidden simulator state. The exposed decision boundary consists of $(\omega_i(\tau_k), \kappa_i(\tau_k), m_i^\Gamma(\cdot \mid \omega_i(\tau_k), \kappa_i(\tau_k)))$, where $\omega_i(\tau_k)=O_i(x_{\tau_k}^{-})$ is the local observation derived from the pre-decision world state, $\kappa_i(\tau_k)$ is the protocol context returned by the interpreter, and $m_i^\Gamma$ is the legality mask induced by the coordination protocol. 
The feasible decision set can be written as
\begin{equation}
\mathcal{A}_i^\Gamma\!\bigl(\omega_i(\tau_k), \kappa_i(\tau_k)\bigr)
=
\left\{
 a \in \mathcal{A}_i
 \;:\;
 m_i^\Gamma\!\bigl(a \mid \omega_i(\tau_k), \kappa_i(\tau_k)\bigr)=1
\right\}.
\end{equation}
Therefore, the same physical state can induce different feasible actions under different $\Gamma$ because the protocol changes who is activated, what information is exposed, and which authority or recovery options are legally available.

\phantomsection\label{app:operational_state_components}
\paragraph{Hidden state, validation, and traces.}
The hidden simulator state remains internal to the event engine and interpreter. Agents never read the full queueing, machine, buffer, transport, blocking, or accounting state directly. These variables influence decisions only through the derived local observation $\omega_i$, protocol context $\kappa_i$, and legal action mask. Returned actions are validated against $\mathcal{A}_i^\Gamma\!\bigl(\omega_i(\tau_k), \kappa_i(\tau_k)\bigr)$ and then committed back to the event engine, which resumes autonomous advancement from the resulting post-decision state. The accounting traces accumulated by the engine support the metric families in Appendix~\ref{app:metric_catalog}.

\section{Coordination Protocol Details}
\label{app:coordination_protocol_details}

This section specifies the implementation-level protocol boundaries corresponding to the four coordination paradigms in Section~\ref{subsec:coordination_paradigms}. In DESBench, each paradigm is implemented as a concrete authority mode that instantiates the coordination protocol $\Gamma$. These authority modes only change how routing authority, local autonomy, rerouting, escalation, and mediated bidding are distributed across the Plant-Area-Cell hierarchy. Let $\rho(\Gamma)=\bigl(q_{\Gamma},u_{\Gamma},p_{\Gamma},h_{\Gamma}\bigr)$ denote the implementation-level authority tuple associated with protocol $\Gamma$, where
\begin{itemize}
    \item $q_{\Gamma}$ denotes the cell-level contract action set;
    \item $u_{\Gamma}$ indicates whether Area-local rerouting is enabled;
    \item $p_{\Gamma}$ indicates whether a failed lower-level commitment is escalated to a mandatory Plant-level reroute path;
    \item $h_{\Gamma}$ indicates whether a mediated bid/award round is enabled.
\end{itemize}

The four DESBench coordination paradigms are shown in Table~\ref{tab:coordination_protocol_switches}.

\textbf{Centralized.}
It is implemented by authority mode \texttt{centralized}. The Plant acts as the direct routing authority and selects the execution cell without requiring an Area-level commitment stage. Cells expose an accept-only contract boundary, so rejection is not part of the effective cell-level action interface. Area-local rerouting and mediated bid rounds are disabled. Because routing authority remains at the Plant, rerouting is handled as direct Plant reselection rather than as upward escalation from a failed local decision.

\textbf{Hierarchical.}
It is implemented by authority mode \texttt{hierarchical}. Routing follows the fixed Plant-Area-Cell chain. The Plant selects an Area, the Area selects a child Cell, and the Cell accepts the assigned commitment under an accept-only contract interface. Area-local rerouting is disabled. If a commitment cannot be settled, the failure propagates upward, and Plant-level reroute is required. This creates a strict layered control flow in which recovery is handled by escalation rather than local sibling substitution.

\textbf{Heterarchical.}
It is implemented by authority mode \texttt{heterarchical\_cnp}. Allocation is mediated through a contract-net-style bid/award mechanism. Cells may participate in a bid round and expose an accept/reject contract boundary. Area-local rerouting is disabled, because conflict resolution is represented through mediated competition.

\textbf{Holonic.}
It is implemented by authority mode \texttt{holonic\_hybrid}. The benchmark preserves the stable Plant-Area-Cell holarchy while granting bounded autonomy to lower levels. Cells expose an accept/reject contract boundary. When a Cell rejects a contract or local infeasibility occurs, the Area may first attempt local rerouting among eligible sibling Cells. Plant fallback is used only after local recovery is exhausted. Mediated bid rounds are disabled, so the protocol emphasizes local self-regulation within a stable holarchy.

\begin{table}[htbp]
\centering
\scriptsize
\setlength{\tabcolsep}{3pt}
\caption{Implementation-level protocol switches for the four coordination paradigms in DESBench.}
\label{tab:coordination_protocol_switches}
\begin{tabular}{@{}
p{0.11\linewidth}
p{0.27\linewidth}
p{0.14\linewidth}
p{0.14\linewidth}
p{0.08\linewidth}
p{0.08\linewidth}
p{0.08\linewidth}@{}}
\toprule
\textbf{Paradigm} &
\textbf{\texttt{mode\_id}} &
\textbf{$q_{\Gamma}$} &
\textbf{$u_{\Gamma}$} &
\textbf{$p_{\Gamma}$} &
\textbf{$h_{\Gamma}$} \\
\midrule
Centralized &
\texttt{centralized} &
\texttt{accept\_only} &
\texttt{False} &
\texttt{False} &
\texttt{False} \\
Hierarchical &
\texttt{hierarchical} &
\texttt{accept\_only} &
\texttt{False} &
\texttt{True} &
\texttt{False} \\
Heterarchical &
\texttt{heterarchical\_cnp} &
\texttt{accept\_reject} &
\texttt{False} &
\texttt{False} &
\texttt{True} \\
Holonic &
\texttt{holonic\_hybrid} &
\texttt{accept\_reject} &
\texttt{True} &
\texttt{False} &
\texttt{False} \\
\bottomrule
\end{tabular}
\setlength{\tabcolsep}{6pt}
\end{table}

\textbf{Action-boundary effects.}
The authority tuple $\rho(\Gamma)$ affects the benchmark interface through the protocol-conditioned legal action set $\mathcal{A}_i^\Gamma\!\bigl(\omega_i(\tau_k), \kappa_i(\tau_k)\bigr).$ For example, when $q_{\Gamma}=\texttt{accept\_only}$, a Cell activation under commitment management exposes only acceptance-compatible contract behavior. When $q_{\Gamma}=\texttt{accept\_reject}$, rejection becomes a legal Cell action and may generate a rejection object in the next materialized protocol set $M_{k+1}$. Similarly, when $u_{\Gamma}=\texttt{True}$, an Area receiving a child rejection may attempt a local reroute before escalation; when $u_{\Gamma}=\texttt{False}$, the same rejection must be settled through the configured escalation or fallback path. When $h_{\Gamma}=\texttt{True}$, the Area-level commitment context includes mediated bid/award semantics, enabling candidate Cells to participate competitively before an allocation is committed.


\subsection{Prompt and Single-Decision Context}
\label{app:prompting_interface}
All framework adapters use the same OpenAI-compatible two-message interaction, so prompt wording is controlled across frameworks and model families. The system instruction only enforces the benchmark contract: the controller must use the current agent-local payload and return exactly one legal action index. The same prompt contract is used for all test cases, and only the runtime payload values change with the current state, activated role, legal actions, and visible context.

\textbf{System instruction.}

\begin{nolinenumbers}
\begin{lstlisting}[style=promptblock]
You are a benchmark controller agent. CRITICAL OUTPUT CONTRACT. Select exactly one legal integer action index from legal_actions. Use only the provided compact agent-local payload, including current context, visible feedback, decision_factors, and bounded working_memory when present. Reply with exactly one JSON object and nothing else. Valid schema: {"action": <index>}. No reasoning. No markdown. No prose.
\end{lstlisting}
\end{nolinenumbers}

\paragraph{Runtime payload construction.}
At each activation, the benchmark first constructs a role-conditioned controller payload from the activated agent's local observation $\omega_i(\tau_k)$ and protocol context $\kappa_i(\tau_k)$. The adapter then compacts this payload into a JSON object with the following recurring fields:
\begin{itemize}
    \item \texttt{agent}, \texttt{role}, and \texttt{kind}: identify the activated controller and decision type;
    \item \texttt{legal\_actions}: the only action indices that can be returned;
    \item \texttt{decision\_factors}: short benchmark-provided guidance for the current decision kind;
    \item \texttt{options}: compact summaries of the currently legal candidates;
    \item \texttt{context}: visible feedback, recent event types, selected job/stage information, and bounded working memory when available.
\end{itemize}

\textbf{Representative single-decision payload from A5C12.}
\begin{nolinenumbers}
\begin{lstlisting}[style=promptblock]
{
  "agent": "plant",
  "role": "plant",
  "kind": "area_selection",
  "legal_actions": [0, 1, 2],
  "decision_factors": [
    "Use current context, visible feedback, and working_memory only.",
    "Prefer feasible actions that reduce congestion and future reroute pressure."
  ],
  "options": [
    {"action": 0, "area_id": 0,
     "summary": {"active_jobs": 0, "backlog_jobs": 10, "energy_kwh": 0.0}},
    {"action": 1, "area_id": 1,
     "summary": {"active_jobs": 0, "backlog_jobs": 10, "energy_kwh": 0.0}},
    {"action": 2, "area_id": 2,
     "summary": {"active_jobs": 0, "backlog_jobs": 10, "energy_kwh": 0.0}}
  ],
  "context": {
    "feedback": {
      "progress": {
        "released_jobs_total": 0,
        "completed_jobs_total": 0,
        "backlog_jobs_total": 10
      },
      "capacity": {
        "admissible_for_new_work": true,
        "busy_machines_total": 0,
        "down_machines_total": 0
      }
    },
    "recent_event_types": ["release_area_selection_window"],
    "selected_job_id": 0,
    "selected_stage_id": 0
  }
}
\end{lstlisting}
\end{nolinenumbers}

Generation uses deterministic decoding with a short output budget, as the benchmark evaluates coordination decisions rather than free-form reasoning. A response is accepted only if it yields a legal integer action from \texttt{legal\_actions}. Otherwise, the runtime discards it, falls back to the contract-safe controller path, and records the decision source for auditing.

\section{Full Metric Catalog}
\label{app:metric_catalog}
DESBench records a broad set of metrics to support both outcome evaluation and deeper analysis of coordination behavior. In Section \ref{subsec:tasks_and_evaluation_dimensions}, we report a minimal reading set chosen for interpretability. This section provides the complete metric catalog.

We group metrics into five categories: \emph{effectiveness}, \emph{constraint alignment}, \emph{coordination efficiency}, \emph{robustness}, and \emph{process-oriented metrics}. The first four correspond to the primary evaluation dimensions used in the main text, capturing schedule quality, adherence to shared operational limits, coordination cost and interaction burden, and resilience under unresolved workload or execution pressure, respectively. The fifth category is reserved for mechanism analysis, trace inspection, and implementation diagnostics, enabling finer-grained investigation of protocol behavior.

\textbf{Effectiveness.}
Table \ref{tab:appendix_metric_effectiveness} describes service quality and task completion outcomes.

\begin{table*}[h]
\centering
\small
\caption{Effectiveness metrics in DESBench.}
\label{tab:appendix_metric_effectiveness}
\begin{tabular}{
  p{0.1\linewidth} 
  p{0.4\linewidth} 
  p{0.4\linewidth}}
\toprule
\textbf{Metric} & \textbf{Meaning} & \textbf{Note} \\
\midrule

$G_{\mathrm{sr}}$ & \texttt{Success rate}: fraction of episodes that terminate successfully. & Strict evaluator success definition: terminated and neither deadlocked nor truncated. \\

$G_{\mathrm{mk}}$ & \texttt{Makespan mean}: mean episode completion time. & Measures realized simulation horizon and is a core effectiveness indicator. \\

$G_{\mathrm{mk}}^{\sigma}$ & \texttt{Makespan std}: standard deviation of episode makespan. & Variability / stability measure across rollouts. \\

$G_{\mathrm{td}}$ & \texttt{Tardiness}: mean total tardiness over completed jobs with due dates. & Counts only completed jobs. \\

$G_{\mathrm{tcd}}$ & \texttt{Tardiness with completion debt}: mean tardiness of completed jobs plus overdue debt from unfinished jobs. & More suitable than $G_{\mathrm{td}}$ when truncation is non‑negligible. \\

$G_{\mathrm{cj}}$ & \texttt{Completed jobs}: mean number of completed jobs per episode. & Raw completion volume. \\

$G_{\mathrm{th}}$ & \texttt{Throughput}: mean value of completed jobs per unit makespan. & Approximate jobs completed per simulation unit time. \\

\bottomrule
\end{tabular}
\end{table*}

\textbf{Constraint Alignment.}
Table \ref{tab:appendix_metric_constraints} quantifies whether the agent completes work while respecting shared resource and budget constraints.

\begin{table*}[h]
\centering
\small
\caption{Constraint‑alignment metrics in DESBench.}
\label{tab:appendix_metric_constraints}
\begin{tabular}{
  p{0.1\linewidth}
  p{0.42\linewidth}
  p{0.38\linewidth}}
\toprule
\textbf{Metric} & \textbf{Meaning} & \textbf{Note} \\
\midrule

$G_{\mathrm{en}}$ & \texttt{Energy}: total energy consumption over an episode, in kWh. & Computed from final cell summaries; core constraint usage indicator. \\

$G_{\mathrm{co}}$ & \texttt{Carbon}: mean total carbon emissions, in kg CO$_2$. & Computed from final cell summaries; reflects sustainability burden. \\

$G_{\mathrm{vc}}$ & \texttt{Violation count}: mean number of recorded overshoot events. & On current structured surfaces, it largely mirrors the frequency of energy and carbon overshoot intervals. \\

$G_{\mathrm{eo}}$ & \texttt{Energy overshoot magnitude}: mean cumulative amount by which energy usage exceeds the cap. & Stronger than pure count metrics, captures overshoot severity. \\

$G_{\mathrm{com}}$ & \texttt{Carbon overshoot magnitude}: mean cumulative carbon overshoot above the cap. & Analogous to energy overshoot but for carbon. \\

$G_{\mathrm{ed}}$ & \texttt{Energy overshoot duration}: mean cumulative time spent above the energy cap. & Time‑above‑cap metric for energy violations. \\

$G_{\mathrm{cod}}$ & \texttt{Carbon overshoot duration}: mean cumulative time spent above the carbon cap. & Time‑above‑cap metric for carbon violations. \\

$G_{\mathrm{eof}}$ & \texttt{Energy overshoot frequency}: mean number of distinct energy‑overshoot intervals. & Frequency‑based diagnostic for overshoot episodes. \\

$G_{\mathrm{cof}}$ & \texttt{Carbon overshoot frequency}: mean number of distinct carbon‑overshoot intervals. & Frequency‑based diagnostic for carbon overshoots. \\

\bottomrule
\end{tabular}
\end{table*}

\textbf{Coordination Efficiency.}
Table \ref{tab:appendix_metric_coordination} quantifies coordination burden, branching complexity, and decision cost. 

\begin{table*}[h]
\centering
\small
\caption{Coordination efficiency metrics in DESBench.}
\label{tab:appendix_metric_coordination}
\begin{tabular}{
  p{0.1\linewidth} 
  p{0.55\linewidth} 
  p{0.25\linewidth}}
\toprule
\textbf{Metric} & \textbf{Meaning} & \textbf{Note} \\
\midrule

$G_{\mathrm{vb}}$ & \texttt{Backlog branching width}: mean number of legal backlog candidates at Plant backlog-selection epochs. & Plant-level branching width. \\

$G_{\mathrm{bc}}$ & \texttt{Backlog decision frequency}: mean number of backlog-selection decision epochs. & Frequency of Plant-level selection events. \\

$G_{\mathrm{pa}}$ & \texttt{Plant routing width}: mean number of legal Areas at Plant routing epochs. & Plant-to-Area routing width. \\

$G_{\mathrm{ac}}$ & \texttt{Area routing width}: mean number of legal Cells at Area decision epochs. & Area-to-Cell branching width. \\

$G_{\mathrm{dw}}$ & \texttt{Local dispatch width}: mean number of legal dispatch options at Cell execution epochs. & Local dispatch branching factor. \\

$G_{\mathrm{ba}}$ & \texttt{Backlog ambiguity rate}: fraction of backlog-selection epochs with multiple legal choices. & Backlog-level ambiguity. \\

$G_{\mathrm{pa}}^{\mathrm{amb}}$ & \texttt{Plant ambiguity rate}: fraction of Plant routing epochs with multiple Areas. & Plant-level ambiguity. \\

$G_{\mathrm{aa}}$ & \texttt{Area ambiguity rate}: fraction of Area decision epochs with multiple Cells. & Area-level ambiguity. \\

$G_{\mathrm{ca}}$ & \texttt{Cell ambiguity rate}: fraction of Cell dispatch epochs with multiple options. & Cell-level ambiguity. \\

$G_{\mathrm{ds}}$ & \texttt{Decision steps}: mean number of decision steps per episode. & High-level coordination burden. \\

$G_{\mathrm{cm}}$ & \texttt{Coordination messages}: mean number of protocol messages exchanged. & Communication cost. \\

$G_{\mathrm{as}}$ & \texttt{Active agent load}: mean cumulative number of agents exposed for decision at each step. & Parallel decision pressure. \\

$G_{\mathrm{wc}}$ & \texttt{runtime\_mean\_seconds}: mean wall-clock runtime, in seconds. & Real execution cost, not simulation time. \\

$G_{\mathrm{llm}}$ & \texttt{Model-driven decision count}: mean number of LLM-driven decisions. & Model-driven control load. \\

$G_{\mathrm{fb}}$ & \texttt{Fallback-contract decisions}: mean number of fallback decisions. & Reliance on rule-based fallback. \\

\bottomrule
\end{tabular}
\end{table*}

\textbf{Robustness.}
Table \ref{tab:appendix_metric_robustness} describes whether an episode converges stably, completely, and without residual work debt. 

\begin{table*}[h]
\centering
\small
\caption{Robustness metrics in DESBench.}
\label{tab:appendix_metric_robustness}
\begin{tabular}{
  p{0.1\linewidth} 
  p{0.55\linewidth} 
  p{0.25\linewidth}}
\toprule
\textbf{Metric} & \textbf{Meaning} & \textbf{Note} \\
\midrule

$G_{\mathrm{dl}}$ & \texttt{Deadlock rate}: fraction of episodes marked as deadlocked. & Hard failure indicator. \\

$G_{\mathrm{tr}}$ & \texttt{Truncation rate}: fraction of episodes marked as truncated. & May arise from horizon limits, deadlock termination, or no-progress stopping. \\

$G_{\mathrm{ut}}$ & \texttt{Unfinished overdue tardiness}: mean overdue debt induced by unfinished jobs at episode end. & Service loss from unfinished workload. \\

$G_{\mathrm{dj}}$ & \texttt{Completion debt (jobs)}: mean number of unfinished jobs at episode end. & Residual job-level debt. \\

$G_{\mathrm{do}}$ & \texttt{Completion debt (operations)}: mean number of remaining operations at episode end. & Residual workload-level debt. \\

\bottomrule
\end{tabular}
\end{table*}

\textbf{Process-oriented Metrics.}
These metrics are not used as the four primary main-text evaluation dimensions. They are more appropriate for trace interpretation, mechanism explanation, and implementation diagnostics. Process-oriented supplementary metrics are shown in Table \ref{tab:appendix_metric_process}. Diagnostic and compatibility metrics are shown in Table \ref{tab:appendix_metric_diagnostic}.

\begin{table*}[h]
\centering
\small
\caption{Process-oriented metrics in DESBench.}
\label{tab:appendix_metric_process}
\begin{tabular}{
  p{0.1\linewidth} 
  p{0.55\linewidth} 
  p{0.25\linewidth}}
\toprule
\textbf{Metric} & \textbf{Meaning} & \textbf{Note} \\
\midrule

$G_{\mathrm{st}}$ & \texttt{Setup time}: mean cumulative setup time within an episode. & Process overhead. \\

$G_{\mathrm{su}}$ & \texttt{Setup energy}: mean extra energy attributable to setup. & Setup cost diagnostic. \\

$G_{\mathrm{tt}}$ & \texttt{Transport time}: mean total transport time. & Logistics burden. \\

$G_{\mathrm{te}}$ & \texttt{Transport energy}: mean transport energy usage. & Logistics cost. \\

$G_{\mathrm{tm}}$ & \texttt{Transport moves}: mean number of transport moves. & Raw logistics activity count. \\

$G_{\mathrm{bt}}$ & \texttt{Blocking time}: mean blocked time induced by downstream or buffer constraints. & Congestion / blocking indicator. \\

$G_{\mathrm{bw}}$ & \texttt{Buffer wait time}: mean waiting time in buffers. & Queueing / waiting indicator. \\

$G_{\mathrm{ae}}$ & \texttt{Arrival events}: mean number of realized arrival events. & Dynamic workload injection count. \\

$G_{\mathrm{aj}}$ & \texttt{Arrived jobs}: mean number of jobs introduced by arrivals. & Dynamic workload scale. \\

$G_{\mathrm{ab}}$ & \texttt{Arrival batch size}: mean jobs per arrival event. & Arrival-shape diagnostic. \\

$G_{\mathrm{bd}}$ & \texttt{Breakdown count}: mean number of machine breakdown events. & Derived from world-event traces. \\

$G_{\mathrm{dt}}$ & \texttt{Downtime}: mean total machine downtime. & Aggregate post-failure unavailability. \\

$G_{\mathrm{rt}}$ & \texttt{Repair time}: mean total repair duration. & Disturbance burden. \\

\bottomrule
\end{tabular}
\end{table*}

\begin{table*}[h]
\centering
\small
\caption{Diagnostic and compatibility metrics in DESBench.}
\label{tab:appendix_metric_diagnostic}
\begin{tabular}{
  p{0.1\linewidth} 
  p{0.55\linewidth} 
  p{0.25\linewidth}}
\toprule
\textbf{Metric} & \textbf{Meaning} & \textbf{Note} \\
\midrule

$G_{\mathrm{vm}}$ & \texttt{Mixed violation magnitude}: legacy mixed violation-magnitude metric. & Aggregates energy and carbon overshoot with inconsistent units. \\

$G_{\mathrm{ic}}$ & \texttt{Inter-cell moves}: mean number of cross-cell moves. & Effectively aliased to transport moves. \\

$G_{\mathrm{ar}}$ & \texttt{Activation records}: mean number of activation-trace records. & Useful for trace volume analysis. \\
$G_{\mathrm{sv}}$ & \texttt{Settlement events}: mean number of settlement-trace events. & Protocol-closure trace indicator. \\

$G_{\mathrm{sc}}$ & \texttt{Settled contracts}: mean number of contracts in settled state at episode end. & Measures contract closure completeness. \\

$G_{\mathrm{oc}}$ & \texttt{Open contracts}: mean number of contracts remaining open at episode end. & Indicates coordination leakage. \\

$G_{\mathrm{uc}}$ & \texttt{Unsettled completed contracts}: mean number of completed but unsettled contracts. & Captures settlement lag. \\

$G_{\mathrm{p1}}$ & \texttt{framework\_pass\_at\_1\_mean}: mean first-pass contract acceptance rate across LLM-backed decisions. & Only available for result bundles that explicitly record pass@1 auditing. \\

$G_{\mathrm{na}}$ & \texttt{No-action events}: mean number of decision points with no legal action available. & Debug/system health indicator. \\

\bottomrule
\end{tabular}
\end{table*}

\section{Implementation Approach}
\label{ia}
This section explains the implementation boundary by which DESBench is instantiated in concrete orchestration frameworks in Section \ref{exp}.

\paragraph{Shared execution contract.}
At each decision epoch, the event interpreter identifies the activated set $\mathcal{U}(\tau_k)$ and constructs one role-conditioned local controller payload for each activated agent. The framework layer receives only these benchmark-visible payloads and must return one legal action per activated controller. Unactivated agents remain idle, and no framework is allowed to read hidden simulator channels outside the controller-facing interface. The shared benchmark runtime therefore remains the single source of truth for world evolution, while the framework layer operates only as a controller orchestrator above it.

\paragraph{Framework realization boundary.}
Both LangGraph \cite{wang2024agent} and AgentScope \cite{gao2024agentscope} are implemented as thin orchestration layers over the same shared event-driven scheduling substrate and controller-facing contract. LangGraph realizes coordination through graph-structured state updates that can carry compact summaries across activations, whereas AgentScope realizes coordination through relay-style message passing with framework-local handoff and observation state. These differences affect how controller-side state is organized inside the framework, but they do not change the benchmark-visible payloads, the legality interface, the event interpreter, or the underlying DES trajectory.


\paragraph{Controller-side statefulness.}
Framework runtimes may maintain lightweight internal state across activations, such as short-horizon summaries of backlog pressure, recent handoff outcomes, or observation histories. This statefulness is intentionally treated as controller-side memory rather than as part of the shared benchmark state. In other words, physical evolution, task progress, and evaluation traces remain in the benchmark runtime, while framework-specific memory affects only how a controller interprets future benchmark-visible payloads. This distinction is important because it permits implementation-side adaptation without allowing frameworks to bypass the common substrate.

\section{Additional Experiments}
\label{ae}
The benchmark defines four reported instances covering different topology scales and difficulty regimes. All experimental results are reported over these instances. Their configurations are summarized in Table~\ref{tab:appendix_instance_catalog}. Each primary table follows the same 15-metric order as the main text. Each logical full-metric matrix is reported as three continued landscape panels: Panel A foregrounds the primary metrics together with closely related outcome and robustness anchors, Panel B reports the broader constraint/resource/process surface, and Panel C reports coordination, audit, and process-diagnostic metrics. Detailed results provided in Tables~\ref{tab:a3c9_cross_primary}--\ref{tab:a5c12_late_full_c}.


\begin{table*}[t]
\centering
\scriptsize
\caption{Benchmark instances. Each reported instance is a stable suite configuration with a fixed topology and route surface.}
\label{tab:appendix_instance_catalog}
\setlength{\tabcolsep}{4pt}
\begin{tabular}{@{}L{0.10\linewidth} L{0.22\linewidth} L{0.23\linewidth} L{0.22\linewidth} L{0.15\linewidth}@{}}
\toprule
\textbf{Instance} & \textbf{Source ID} & \textbf{Topology and route surface} & \textbf{Profile and purpose} & \textbf{Failure and budget} \\
\midrule
\texttt{A3C9-1}
& \path{intercell_a3c9_cross_area_tradeoff}
& \path{tree_a3_c9}; three Areas and nine Cells; three-stage \path{a3c9_canonical_non_degenerate_v1} route grammar
& Cross-area tradeoff; max in-flight jobs $4$; backlog top-$3$; transport multiplier $1.15$; inbound cap $1$; smaller-topology transfer and cross-area routing-regret test.
& Weibull-aging failures from the provided profile metadata; standard R0 WIP, energy, and carbon case budgets \\
\texttt{A5C12-1}
& \path{intercell_a5c12_branch_pressure}
& \path{tree_a5_c12}; five Areas and twelve Cells with a 4--3--2--2--1 asymmetric partition; four-stage \path{a5c12_asymmetric_long_horizon_v1} route grammar
& Branch-pressure anchor; max in-flight jobs $4$; backlog top-$4$; transport multiplier $1.15$; default inbound cap $1$; later weak cells $(3,6,8,10,11)$ are slowed.
& Exponential nominal failures; A5C12 R0 WIP, energy, and carbon budgets fixed across protocols. \\
\texttt{A5C12-2}
& \path{intercell_a5c12_cluster_pull}
& \path{tree_a5_c12}; five Areas and twelve Cells with a 4--3--2--2--1 asymmetric partition; four-stage \path{a5c12_asymmetric_long_horizon_v1} route grammar
& Strong-cluster-pull / greedy routing lure; max in-flight jobs $4$; backlog top-$4$; transport multiplier $1.20$; default inbound cap $1$; cells $(0,1,4)$ are faster while $(7,9,11)$ are slower.
& Weibull-aging failures; same A5C12 R0 budget-accounting surface as \texttt{A5C12-1}. \\
\texttt{A5C12-3}
& \path{intercell_a5c12_late_commit}
& \path{tree_a5_c12}; five Areas and twelve Cells with a 4--3--2--2--1 asymmetric partition; four-stage \path{a5c12_asymmetric_long_horizon_v1} route grammar
& Late-commitment / long-horizon commitment regret; max in-flight jobs $4$; backlog top-$4$; transport multiplier $1.25$; default inbound cap $1$; late-stage weak cells $(2,6,8,10,11)$ are slowed.
& Load-dependent failures; same A5C12 R0 budget-accounting surface, emphasizing long-horizon commitment cost. \\
\bottomrule
\end{tabular}
\end{table*}


\begin{table*}[htbp]
\centering
\setlength{\tabcolsep}{3pt}
\renewcommand{\arraystretch}{1.08}
\caption{Primary 15-metric matrix for \texttt{A3C9-1}.}
\label{tab:a3c9_cross_primary}
\resizebox{\textwidth}{!}{%
\begin{tabular}{lll*{15}{c}}
\toprule
\multirow{3}{*}{\textbf{Paradigm}} & \multirow{3}{*}{\textbf{Framework}} & \multirow{3}{*}{\textbf{LLM}} & \multicolumn{15}{c}{\textbf{Evaluation Metrics}} \\
\cmidrule(lr){4-18}
& & & \multicolumn{3}{c}{$\mathcal{G}_{\mathrm{eff}}$} & \multicolumn{4}{c}{$\mathcal{G}_{\mathrm{con}}$} & \multicolumn{4}{c}{$\mathcal{G}_{\mathrm{coord}}$} & \multicolumn{4}{c}{$\mathcal{G}_{\mathrm{rob}}$} \\
\cmidrule(lr){4-6}\cmidrule(lr){7-10}\cmidrule(lr){11-14}\cmidrule(lr){15-18}
& & & $G_{\mathrm{mk}} \downarrow$ & $G_{\mathrm{td}} \downarrow$ & $G_{\mathrm{cj}} \uparrow$ & $G_{\mathrm{vc}} \downarrow$ & $G_{\mathrm{eo}} \downarrow$ & $G_{\mathrm{ed}} \downarrow$ & $G_{\mathrm{en}} \downarrow$ & $G_{\mathrm{ds}} \downarrow$ & $G_{\mathrm{cm}} \downarrow$ & $G_{\mathrm{fb}} \downarrow$ & $G_{\mathrm{llm}} \downarrow$ & $G_{\mathrm{tr}} \downarrow$ & $G_{\mathrm{dj}} \downarrow$ & $G_{\mathrm{ut}} \downarrow$ & $G_{\mathrm{do}} \downarrow$ \\
\midrule
\multirow{6}{*}{$\Gamma^{\mathrm{cen}}$} & \multirow{3}{*}{LangGraph} & GPT-5.4 & 54.7 & 116.2 & 10.0 & 22.2 & 3277.2 & 5.0 & 6562.7 & 90.0 & 90.0 & 0.00 & 90.0 & 0.00 & 0.00 & 0.00 & 0.00 \\
  &   & Gemini-3-Flash & 55.2 & 107.5 & 10.0 & 20.6 & 3542.1 & 6.1 & 6620.9 & 90.0 & 90.0 & 1.6 & 88.4 & 0.00 & 0.00 & 0.00 & 0.00 \\
  &   & Qwen-3.5 & 54.7 & 115.1 & 10.0 & 23.2 & 2962.8 & 6.0 & 6554.1 & 90.0 & 90.0 & 0.00 & 90.0 & 0.00 & 0.00 & 0.00 & 0.00 \\
\cmidrule(lr){2-18}
  & \multirow{3}{*}{AgentScope} & GPT-5.4 & 55.2 & 121.0 & 10.0 & 21.8 & 3363.8 & 5.5 & 6622.2 & 90.0 & 90.0 & 0.00 & 90.0 & 0.00 & 0.00 & 0.00 & 0.00 \\
  &   & Gemini-3-Flash & 55.0 & 106.7 & 10.0 & 20.6 & 3564.7 & 6.0 & 6602.4 & 90.0 & 90.0 & 1.8 & 88.2 & 0.00 & 0.00 & 0.00 & 0.00 \\
\bottomrule
\end{tabular}%
}
\end{table*}

\begin{sidewaystable*}[p]
\centering
\scriptsize
\setlength{\tabcolsep}{1.9pt}
\renewcommand{\arraystretch}{1.05}
\caption{Logical full-metric matrix for \texttt{A3C9-1} (Panel A)}
\label{tab:a3c9_cross_full}
\resizebox{\textheight}{!}{%
\begin{tabular}{@{}lll*{18}{c}@{}}
\toprule
\multirow{2}{*}{\textbf{Paradigm}} & \multirow{2}{*}{\textbf{Framework}} & \multirow{2}{*}{\textbf{LLM}} & \multicolumn{18}{c}{\textbf{Panel A: Primary Outcomes + Robustness Anchors}} \\\\
\cmidrule(lr){4-21}
 & & & \rotatebox{90}{$G_{\mathrm{mk}}$} & \rotatebox{90}{$G_{\mathrm{mk}}^{\sigma}$} & \rotatebox{90}{$G_{\mathrm{td}}$} & \rotatebox{90}{$G_{\mathrm{tcd}}$} & \rotatebox{90}{$G_{\mathrm{cj}}$} & \rotatebox{90}{$G_{\mathrm{th}}$} & \rotatebox{90}{$G_{\mathrm{sr}}$} & \rotatebox{90}{$G_{\mathrm{dl}}$} & \rotatebox{90}{$G_{\mathrm{tr}}$} & \rotatebox{90}{$G_{\mathrm{ut}}$} & \rotatebox{90}{$G_{\mathrm{dj}}$} & \rotatebox{90}{$G_{\mathrm{do}}$} & \rotatebox{90}{$G_{\mathrm{en}}$} & \rotatebox{90}{$G_{\mathrm{vc}}$} & \rotatebox{90}{$G_{\mathrm{eo}}$} & \rotatebox{90}{$G_{\mathrm{ed}}$} & \rotatebox{90}{$G_{\mathrm{ds}}$} & \rotatebox{90}{$G_{\mathrm{cm}}$} \\\\
\midrule
$\Gamma^{\mathrm{cen}}$ & LangGraph & GPT-5.4 & 54.7 & 1.57 & 116.2 & 116.2 & 10.0 & 0.183 & 1.00 & 0.000 & 0.000 & 0.000 & 0.000 & 0.000 & 6562.7 & 22.2 & 3277.2 & 5.00 & 90.0 & 90.0 \\
$\Gamma^{\mathrm{cen}}$ & LangGraph & Gemini-3-Flash & 55.2 & 1.08 & 107.5 & 107.5 & 10.0 & 0.181 & 1.00 & 0.000 & 0.000 & 0.000 & 0.000 & 0.000 & 6620.9 & 20.6 & 3542.1 & 6.13 & 90.0 & 90.0 \\
$\Gamma^{\mathrm{cen}}$ & LangGraph & Qwen-3.5 & 54.7 & 2.05 & 115.1 & 115.1 & 10.0 & 0.183 & 1.00 & 0.000 & 0.000 & 0.000 & 0.000 & 0.000 & 6554.1 & 23.2 & 2962.8 & 6.01 & 90.0 & 90.0 \\
$\Gamma^{\mathrm{cen}}$ & AgentScope & GPT-5.4 & 55.2 & 2.55 & 121.0 & 121.0 & 10.0 & 0.182 & 1.00 & 0.000 & 0.000 & 0.000 & 0.000 & 0.000 & 6622.2 & 21.8 & 3363.8 & 5.49 & 90.0 & 90.0 \\
\bottomrule
\end{tabular}%
}
\end{sidewaystable*}

\begin{sidewaystable*}[p]
\centering
\scriptsize
\setlength{\tabcolsep}{1.55pt}
\renewcommand{\arraystretch}{1.05}
\caption[]{Logical full-metric matrix for \texttt{A3C9-1} (Panel B).}
\label{tab:a3c9_cross_full_b}
\resizebox{\textheight}{!}{%
\begin{tabular}{@{}lll*{20}{c}@{}}
\toprule
\multirow{2}{*}{\textbf{Paradigm}} & \multirow{2}{*}{\textbf{Framework}} & \multirow{2}{*}{\textbf{LLM}} & \multicolumn{20}{c}{\textbf{Panel B: Constraint / Resource / Process}} \\\\
\cmidrule(lr){4-23}
 & & & \rotatebox{90}{$G_{\mathrm{co}}$} & \rotatebox{90}{$G_{\mathrm{vm}}$} & \rotatebox{90}{$G_{\mathrm{com}}$} & \rotatebox{90}{$G_{\mathrm{cod}}$} & \rotatebox{90}{$G_{\mathrm{eof}}$} & \rotatebox{90}{$G_{\mathrm{cof}}$} & \rotatebox{90}{$G_{\mathrm{st}}$} & \rotatebox{90}{$G_{\mathrm{su}}$} & \rotatebox{90}{$G_{\mathrm{tt}}$} & \rotatebox{90}{$G_{\mathrm{te}}$} & \rotatebox{90}{$G_{\mathrm{tm}}$} & \rotatebox{90}{$G_{\mathrm{ic}}$} & \rotatebox{90}{$G_{\mathrm{bt}}$} & \rotatebox{90}{$G_{\mathrm{bw}}$} & \rotatebox{90}{$G_{\mathrm{ae}}$} & \rotatebox{90}{$G_{\mathrm{aj}}$} & \rotatebox{90}{$G_{\mathrm{ab}}$} & \rotatebox{90}{$G_{\mathrm{bd}}$} & \rotatebox{90}{$G_{\mathrm{dt}}$} & \rotatebox{90}{$G_{\mathrm{rt}}$} \\\\
\midrule
$\Gamma^{\mathrm{cen}}$ & LangGraph & GPT-5.4 & 3937.6 & 5243.5 & 1966.3 & 5.00 & 11.1 & 11.1 & 5.32 & 31.9 & 49.2 & 9.84 & 30.0 & 30.0 & 0.000 & 0.000 & 4.00 & 7.00 & 1.75 & 0.100 & 0.011 & 0.011 \\
$\Gamma^{\mathrm{cen}}$ & LangGraph & Gemini-3-Flash & 3972.5 & 5667.4 & 2125.3 & 6.13 & 10.3 & 10.3 & 5.04 & 30.3 & 48.9 & 9.78 & 30.0 & 30.0 & 0.000 & 0.000 & 4.00 & 7.00 & 1.75 & 0.000 & 0.000 & 0.000 \\
$\Gamma^{\mathrm{cen}}$ & LangGraph & Qwen-3.5 & 3932.4 & 4740.5 & 1777.7 & 6.01 & 11.6 & 11.6 & 4.15 & 24.9 & 48.6 & 9.73 & 30.0 & 30.0 & 0.000 & 0.000 & 4.00 & 7.00 & 1.75 & 0.000 & 0.000 & 0.000 \\
$\Gamma^{\mathrm{cen}}$ & AgentScope & GPT-5.4 & 3973.3 & 5382.1 & 2018.3 & 5.49 & 10.9 & 10.9 & 4.76 & 28.6 & 49.5 & 9.89 & 30.0 & 30.0 & 0.000 & 0.000 & 4.00 & 7.00 & 1.75 & 0.100 & 0.925 & 0.925 \\
\bottomrule
\end{tabular}%
}
\end{sidewaystable*}

\begin{sidewaystable*}[p]
\centering
\scriptsize
\setlength{\tabcolsep}{1.7pt}
\renewcommand{\arraystretch}{1.05}
\caption[]{Logical full-metric matrix for \texttt{A3C9-1} ((Panel C).}
\label{tab:a3c9_cross_full_c}
\resizebox{\textheight}{!}{%
\begin{tabular}{@{}lll*{19}{c}@{}}
\toprule
\multirow{2}{*}{\textbf{Paradigm}} & \multirow{2}{*}{\textbf{Framework}} & \multirow{2}{*}{\textbf{LLM}} & \multicolumn{19}{c}{\textbf{Panel C: Coordination / Audit / Diagnostics}} \\\\
\cmidrule(lr){4-22}
 & & & \rotatebox{90}{$G_{\mathrm{vb}}$} & \rotatebox{90}{$G_{\mathrm{bc}}$} & \rotatebox{90}{$G_{\mathrm{pa}}$} & \rotatebox{90}{$G_{\mathrm{ac}}$} & \rotatebox{90}{$G_{\mathrm{dw}}$} & \rotatebox{90}{$G_{\mathrm{ba}}$} & \rotatebox{90}{$G_{\mathrm{pa}}^{\mathrm{amb}}$} & \rotatebox{90}{$G_{\mathrm{aa}}$} & \rotatebox{90}{$G_{\mathrm{ca}}$} & \rotatebox{90}{$G_{\mathrm{as}}$} & \rotatebox{90}{$G_{\mathrm{wc}}$} & \rotatebox{90}{$G_{\mathrm{ar}}$} & \rotatebox{90}{$G_{\mathrm{sv}}$} & \rotatebox{90}{$G_{\mathrm{sc}}$} & \rotatebox{90}{$G_{\mathrm{oc}}$} & \rotatebox{90}{$G_{\mathrm{uc}}$} & \rotatebox{90}{$G_{\mathrm{llm}}$} & \rotatebox{90}{$G_{\mathrm{fb}}$} & \rotatebox{90}{$G_{\mathrm{na}}$} \\\\
\midrule
$\Gamma^{\mathrm{cen}}$ & LangGraph & GPT-5.4 & 2.10 & 30.0 & 0.000 & 0.000 & 2.10 & 0.710 & 0.000 & 0.000 & 0.733 & 90.0 & 1095.9 & 90.0 & 120.0 & 10.0 & 20.0 & 0.000 & 90.0 & 0.000 & 0.000 \\
$\Gamma^{\mathrm{cen}}$ & LangGraph & Gemini-3-Flash & 2.17 & 30.0 & 0.000 & 0.000 & 2.10 & 0.677 & 0.000 & 0.000 & 0.733 & 90.0 & 417.0 & 90.0 & 120.0 & 10.0 & 20.0 & 0.000 & 88.4 & 1.60 & 0.000 \\
$\Gamma^{\mathrm{cen}}$ & LangGraph & Qwen-3.5 & 2.17 & 30.0 & 0.000 & 0.000 & 2.10 & 0.777 & 0.000 & 0.000 & 0.733 & 90.0 & 108.3 & 90.0 & 120.0 & 10.0 & 20.0 & 0.000 & 90.0 & 0.000 & 0.000 \\
$\Gamma^{\mathrm{cen}}$ & AgentScope & GPT-5.4 & 2.09 & 30.0 & 0.000 & 0.000 & 2.10 & 0.703 & 0.000 & 0.000 & 0.733 & 90.0 & 1094.3 & 90.0 & 120.0 & 10.0 & 20.0 & 0.000 & 90.0 & 0.000 & 0.000 \\
\bottomrule
\end{tabular}%
}
\end{sidewaystable*}

\begin{table*}[htbp]
\centering
\setlength{\tabcolsep}{3pt}
\renewcommand{\arraystretch}{1.08}
\caption{Primary 15-metric matrix for \texttt{A5C12-1}.}
\label{tab:a5c12_branch_primary}
\resizebox{\textwidth}{!}{%
\begin{tabular}{lll*{15}{c}}
\toprule
\multirow{3}{*}{\textbf{Paradigm}} & \multirow{3}{*}{\textbf{Framework}} & \multirow{3}{*}{\textbf{LLM}} & \multicolumn{15}{c}{\textbf{Evaluation Metrics}} \\
\cmidrule(lr){4-18}
& & & \multicolumn{3}{c}{$\mathcal{G}_{\mathrm{eff}}$} & \multicolumn{4}{c}{$\mathcal{G}_{\mathrm{con}}$} & \multicolumn{4}{c}{$\mathcal{G}_{\mathrm{coord}}$} & \multicolumn{4}{c}{$\mathcal{G}_{\mathrm{rob}}$} \\
\cmidrule(lr){4-6}\cmidrule(lr){7-10}\cmidrule(lr){11-14}\cmidrule(lr){15-18}
& & & $G_{\mathrm{mk}} \downarrow$ & $G_{\mathrm{td}}\downarrow$ & $G_{\mathrm{cj}}\uparrow$ & $G_{\mathrm{vc}}\downarrow$ & $G_{\mathrm{eo}}\downarrow$ & $G_{\mathrm{ed}}\downarrow$ & $G_{\mathrm{en}}\downarrow$ & $G_{\mathrm{ds}}\downarrow$ & $G_{\mathrm{cm}}\downarrow$ & $G_{\mathrm{fb}}\downarrow$ & $G_{\mathrm{llm}}\downarrow$ & $G_{\mathrm{tr}}\downarrow$ & $G_{\mathrm{dj}}\downarrow$ & $G_{\mathrm{ut}}\downarrow$ & $G_{\mathrm{do}}\downarrow$ \\
\midrule
\multirow{6}{*}{$\Gamma^{\mathrm{cen}}$} & \multirow{3}{*}{LangGraph} & GPT-5.4 & 78.2 & 242.3 & 10.0 & 32.4 & 12636.8 & 10.7 & 12255.4 & 120.0 & 120.0 & 0.2 & 119.8 & 0.00 & 0.0 & 0.0 & 0.0 \\
 &  & Gemini-3-Flash & 73.9 & 230.2 & 10.0 & 31.2 & 10389.3 & 8.0 & 11621.4 & 120.0 & 120.0 & 2.2 & 117.8 & 0.00 & 0.0 & 0.0 & 0.0 \\
 &  & Qwen-3.5 & 68.8 & 211.1 & 9.4 & 27.6 & 8862.4 & 6.2 & 10851.9 & 114.9 & 114.9 & 0.0 & 114.9 & 0.10 & 0.6 & 6.1 & 2.0 \\
\cmidrule(lr){2-18}
 & \multirow{3}{*}{AgentScope} & GPT-5.4 & 77.4 & 253.2 & 10.0 & 33.2 & 13662.8 & 9.3 & 12135.3 & 120.0 & 120.0 & 0.1 & 119.9 & 0.00 & 0.0 & 0.0 & 0.0 \\
 &  & Gemini-3-Flash & 73.9 & 230.2 & 10.0 & 31.2 & 10388.9 & 8.0 & 11621.1 & 120.0 & 120.0 & 2.2 & 117.8 & 0.00 & 0.0 & 0.0 & 0.0 \\
\bottomrule
\end{tabular}%
}
\end{table*}

\begin{sidewaystable*}[p]
\centering
\scriptsize
\setlength{\tabcolsep}{1.9pt}
\renewcommand{\arraystretch}{1.05}
\caption{Logical full-metric matrix for \texttt{A5C12-1} (Panel A).}
\label{tab:a5c12_branch_full}
\resizebox{\textheight}{!}{%
\begin{tabular}{@{}lll*{18}{c}@{}}
\toprule
\multirow{2}{*}{\textbf{Paradigm}} & \multirow{2}{*}{\textbf{Framework}} & \multirow{2}{*}{\textbf{LLM}} & \multicolumn{18}{c}{\textbf{Panel A: Primary Outcomes + Robustness Anchors}} \\\\
\cmidrule(lr){4-21}
 & & & \rotatebox{90}{$G_{\mathrm{mk}}$} & \rotatebox{90}{$G_{\mathrm{mk}}^{\sigma}$} & \rotatebox{90}{$G_{\mathrm{td}}$} & \rotatebox{90}{$G_{\mathrm{tcd}}$} & \rotatebox{90}{$G_{\mathrm{cj}}$} & \rotatebox{90}{$G_{\mathrm{th}}$} & \rotatebox{90}{$G_{\mathrm{sr}}$} & \rotatebox{90}{$G_{\mathrm{dl}}$} & \rotatebox{90}{$G_{\mathrm{tr}}$} & \rotatebox{90}{$G_{\mathrm{ut}}$} & \rotatebox{90}{$G_{\mathrm{dj}}$} & \rotatebox{90}{$G_{\mathrm{do}}$} & \rotatebox{90}{$G_{\mathrm{en}}$} & \rotatebox{90}{$G_{\mathrm{vc}}$} & \rotatebox{90}{$G_{\mathrm{eo}}$} & \rotatebox{90}{$G_{\mathrm{ed}}$} & \rotatebox{90}{$G_{\mathrm{ds}}$} & \rotatebox{90}{$G_{\mathrm{cm}}$} \\\\
\midrule
$\Gamma^{\mathrm{cen}}$ & LangGraph & GPT-5.4 & 78.2 & 4.39 & 242.3 & 242.3 & 10.0 & 0.128 & 1.00 & 0.000 & 0.000 & 0.000 & 0.000 & 0.000 & 12255.4 & 32.4 & 12636.8 & 10.7 & 120.0 & 120.0 \\
$\Gamma^{\mathrm{cen}}$ & LangGraph & Gemini-3-Flash & 73.9 & 2.18 & 230.2 & 230.2 & 10.0 & 0.135 & 1.00 & 0.000 & 0.000 & 0.000 & 0.000 & 0.000 & 11621.4 & 31.2 & 10389.3 & 8.04 & 120.0 & 120.0 \\
$\Gamma^{\mathrm{cen}}$ & LangGraph & Qwen-3.5 & 68.8 & 10.6 & 211.1 & 217.2 & 9.40 & 0.135 & 0.900 & 0.000 & 0.100 & 6.10 & 0.600 & 2.00 & 10851.9 & 27.6 & 8862.4 & 6.20 & 114.9 & 114.9 \\
$\Gamma^{\mathrm{cen}}$ & AgentScope & GPT-5.4 & 77.4 & 4.41 & 253.2 & 253.2 & 10.0 & 0.130 & 1.00 & 0.000 & 0.000 & 0.000 & 0.000 & 0.000 & 12135.3 & 33.2 & 13662.8 & 9.28 & 120.0 & 120.0 \\
\bottomrule
\end{tabular}%
}
\end{sidewaystable*}

\begin{sidewaystable*}[p]
\centering
\scriptsize
\setlength{\tabcolsep}{1.55pt}
\renewcommand{\arraystretch}{1.05}
\caption[]{Logical full-metric matrix for \texttt{A5C12-1} (Panel B).}
\label{tab:a5c12_branch_full_b}
\resizebox{\textheight}{!}{%
\begin{tabular}{@{}lll*{20}{c}@{}}
\toprule
\multirow{2}{*}{\textbf{Paradigm}} & \multirow{2}{*}{\textbf{Framework}} & \multirow{2}{*}{\textbf{LLM}} & \multicolumn{20}{c}{\textbf{Panel B: Constraint / Resource / Process}} \\\\
\cmidrule(lr){4-23}
 & & & \rotatebox{90}{$G_{\mathrm{co}}$} & \rotatebox{90}{$G_{\mathrm{vm}}$} & \rotatebox{90}{$G_{\mathrm{com}}$} & \rotatebox{90}{$G_{\mathrm{cod}}$} & \rotatebox{90}{$G_{\mathrm{eof}}$} & \rotatebox{90}{$G_{\mathrm{cof}}$} & \rotatebox{90}{$G_{\mathrm{st}}$} & \rotatebox{90}{$G_{\mathrm{su}}$} & \rotatebox{90}{$G_{\mathrm{tt}}$} & \rotatebox{90}{$G_{\mathrm{te}}$} & \rotatebox{90}{$G_{\mathrm{tm}}$} & \rotatebox{90}{$G_{\mathrm{ic}}$} & \rotatebox{90}{$G_{\mathrm{bt}}$} & \rotatebox{90}{$G_{\mathrm{bw}}$} & \rotatebox{90}{$G_{\mathrm{ae}}$} & \rotatebox{90}{$G_{\mathrm{aj}}$} & \rotatebox{90}{$G_{\mathrm{ab}}$} & \rotatebox{90}{$G_{\mathrm{bd}}$} & \rotatebox{90}{$G_{\mathrm{dt}}$} & \rotatebox{90}{$G_{\mathrm{rt}}$} \\\\
\midrule
$\Gamma^{\mathrm{cen}}$ & LangGraph & GPT-5.4 & 7353.2 & 20218.8 & 7582.1 & 10.7 & 16.2 & 16.2 & 9.45 & 56.7 & 66.0 & 13.2 & 40.0 & 40.0 & 0.000 & 0.000 & 4.00 & 7.00 & 1.75 & 0.500 & 1.68 & 1.68 \\
$\Gamma^{\mathrm{cen}}$ & LangGraph & Gemini-3-Flash & 6972.9 & 16622.8 & 6233.6 & 8.04 & 15.6 & 15.6 & 9.50 & 57.0 & 65.0 & 13.0 & 40.0 & 40.0 & 0.000 & 0.000 & 4.00 & 7.00 & 1.75 & 0.500 & 3.16 & 3.16 \\
$\Gamma^{\mathrm{cen}}$ & LangGraph & Qwen-3.5 & 6511.1 & 14179.9 & 5317.4 & 6.20 & 13.8 & 13.8 & 9.27 & 55.6 & 62.2 & 12.4 & 38.4 & 38.4 & 0.000 & 0.000 & 4.00 & 7.00 & 1.75 & 0.300 & 0.727 & 0.727 \\
$\Gamma^{\mathrm{cen}}$ & AgentScope & GPT-5.4 & 7281.2 & 21860.4 & 8197.7 & 9.28 & 16.6 & 16.6 & 8.88 & 53.3 & 66.2 & 13.2 & 40.0 & 40.0 & 0.000 & 0.000 & 4.00 & 7.00 & 1.75 & 0.500 & 3.69 & 3.69 \\
\bottomrule
\end{tabular}%
}
\end{sidewaystable*}

\begin{sidewaystable*}[p]
\centering
\scriptsize
\setlength{\tabcolsep}{1.7pt}
\renewcommand{\arraystretch}{1.05}
\caption[]{Logical full-metric matrix for \texttt{A5C12-1} (Panel C).}
\label{tab:a5c12_branch_full_c}
\resizebox{\textheight}{!}{%
\begin{tabular}{@{}lll*{20}{c}@{}}
\toprule
\multirow{2}{*}{\textbf{Paradigm}} & \multirow{2}{*}{\textbf{Framework}} & \multirow{2}{*}{\textbf{LLM}} & \multicolumn{20}{c}{\textbf{Panel C: Coordination / Audit / Diagnostics}} \\\\
\cmidrule(lr){4-23}
 & & & \rotatebox{90}{$G_{\mathrm{vb}}$} & \rotatebox{90}{$G_{\mathrm{bc}}$} & \rotatebox{90}{$G_{\mathrm{pa}}$} & \rotatebox{90}{$G_{\mathrm{ac}}$} & \rotatebox{90}{$G_{\mathrm{dw}}$} & \rotatebox{90}{$G_{\mathrm{ba}}$} & \rotatebox{90}{$G_{\mathrm{pa}}^{\mathrm{amb}}$} & \rotatebox{90}{$G_{\mathrm{aa}}$} & \rotatebox{90}{$G_{\mathrm{ca}}$} & \rotatebox{90}{$G_{\mathrm{as}}$} & \rotatebox{90}{$G_{\mathrm{wc}}$} & \rotatebox{90}{$G_{\mathrm{ar}}$} & \rotatebox{90}{$G_{\mathrm{sv}}$} & \rotatebox{90}{$G_{\mathrm{sc}}$} & \rotatebox{90}{$G_{\mathrm{oc}}$} & \rotatebox{90}{$G_{\mathrm{uc}}$} & \rotatebox{90}{$G_{\mathrm{llm}}$} & \rotatebox{90}{$G_{\mathrm{fb}}$} & \rotatebox{90}{$G_{\mathrm{na}}$} & \rotatebox{90}{$G_{\mathrm{p1}}$} \\\\
\midrule
$\Gamma^{\mathrm{cen}}$ & LangGraph & GPT-5.4 & 2.00 & 40.0 & 0.000 & 0.000 & 2.00 & 0.555 & 0.000 & 0.000 & 0.700 & 120.0 & 146.7 & 120.0 & 160.0 & 10.0 & 30.0 & 0.000 & 119.8 & 0.200 & 0.000 & 0.998 \\
$\Gamma^{\mathrm{cen}}$ & LangGraph & Gemini-3-Flash & 1.96 & 40.0 & 0.000 & 0.000 & 2.02 & 0.515 & 0.000 & 0.000 & 0.698 & 120.0 & 257.3 & 120.0 & 160.0 & 10.0 & 30.0 & 0.000 & 117.8 & 2.20 & 0.000 & -- \\
$\Gamma^{\mathrm{cen}}$ & LangGraph & Qwen-3.5 & 1.91 & 38.4 & 0.000 & 0.000 & 2.01 & 0.487 & 0.000 & 0.000 & 0.701 & 114.9 & 191.6 & 114.9 & 152.8 & 9.40 & 29.0 & 0.000 & 114.9 & 0.000 & 0.000 & 1.00 \\
$\Gamma^{\mathrm{cen}}$ & AgentScope & GPT-5.4 & 2.00 & 40.0 & 0.000 & 0.000 & 2.02 & 0.545 & 0.000 & 0.000 & 0.700 & 120.0 & 355.8 & 120.0 & 160.0 & 10.0 & 30.0 & 0.000 & 119.9 & 0.100 & 0.000 & 0.999 \\
\bottomrule
\end{tabular}%
}
\end{sidewaystable*}

\begin{table*}[htbp]
\centering
\setlength{\tabcolsep}{3pt}
\renewcommand{\arraystretch}{1.08}
\caption{Primary 15-metric matrix for \texttt{A5C12-2}.}
\label{tab:a5c12_cluster_primary}
\resizebox{\textwidth}{!}{%
\begin{tabular}{lll*{15}{c}}
\toprule
\multirow{3}{*}{\textbf{Paradigm}} & \multirow{3}{*}{\textbf{Framework}} & \multirow{3}{*}{\textbf{LLM}} & \multicolumn{15}{c}{\textbf{Evaluation Metrics}} \\
\cmidrule(lr){4-18}
& & & \multicolumn{3}{c}{$\mathcal{G}_{\mathrm{eff}}$} & \multicolumn{4}{c}{$\mathcal{G}_{\mathrm{con}}$} & \multicolumn{4}{c}{$\mathcal{G}_{\mathrm{coord}}$} & \multicolumn{4}{c}{$\mathcal{G}_{\mathrm{rob}}$} \\
\cmidrule(lr){4-6}\cmidrule(lr){7-10}\cmidrule(lr){11-14}\cmidrule(lr){15-18}
& & & $G_{\mathrm{mk}} \downarrow$ & $G_{\mathrm{td}}\downarrow$ & $G_{\mathrm{cj}}\uparrow$ & $G_{\mathrm{vc}}\downarrow$ & $G_{\mathrm{eo}}\downarrow$ & $G_{\mathrm{ed}}\downarrow$ & $G_{\mathrm{en}}\downarrow$ & $G_{\mathrm{ds}}\downarrow$ & $G_{\mathrm{cm}}\downarrow$ & $G_{\mathrm{fb}}\downarrow$ & $G_{\mathrm{llm}}\downarrow$ & $G_{\mathrm{tr}}\downarrow$ & $G_{\mathrm{dj}}\downarrow$ & $G_{\mathrm{ut}}\downarrow$ & $G_{\mathrm{do}}\downarrow$ \\
\midrule
\multirow{6}{*}{$\Gamma^{\mathrm{cen}}$} & \multirow{3}{*}{LangGraph} & GPT-5.4 & 78.3 & 261.2 & 10.0 & 32.6 & 13454.4 & 9.4 & 12238.7 & 120.0 & 120.0 & 0.0 & 120.0 & 0.00 & 0.0 & 0.0 & 0.0 \\
 &  & Gemini-3-Flash & 72.9 & 248.4 & 10.0 & 32.2 & 10924.9 & 4.8 & 11453.1 & 120.0 & 120.0 & 0.0 & 120.0 & 0.00 & 0.0 & 0.0 & 0.0 \\
 &  & Qwen-3.5 & 74.0 & 253.9 & 10.0 & 32.0 & 11571.3 & 6.1 & 11625.5 & 120.0 & 120.0 & 0.0 & 120.0 & 0.00 & 0.0 & 0.0 & 0.0 \\
\cmidrule(lr){2-18}
 & \multirow{3}{*}{AgentScope} & GPT-5.4 & 68.6 & 208.3 & 8.3 & 26.4 & 9912.3 & 7.6 & 10722.2 & 106.2 & 105.9 & 0.1 & 106.1 & 0.20 & 1.7 & 10.1 & 5.3 \\
 &  & Gemini-3-Flash & 72.9 & 248.4 & 10.0 & 32.2 & 10924.9 & 4.8 & 11453.1 & 120.0 & 120.0 & 0.0 & 120.0 & 0.00 & 0.0 & 0.0 & 0.0 \\
\bottomrule
\end{tabular}%
}
\end{table*}

\begin{sidewaystable*}[p]
\centering
\scriptsize
\setlength{\tabcolsep}{1.9pt}
\renewcommand{\arraystretch}{1.05}
\caption{Logical full-metric matrix for \texttt{A5C12-2} (Panel A).}
\label{tab:a5c12_cluster_full}
\resizebox{\textheight}{!}{%
\begin{tabular}{@{}lll*{18}{c}@{}}
\toprule
\multirow{2}{*}{\textbf{Paradigm}} & \multirow{2}{*}{\textbf{Framework}} & \multirow{2}{*}{\textbf{LLM}} & \multicolumn{18}{c}{\textbf{Panel A: Primary Outcomes + Robustness Anchors}} \\\\
\cmidrule(lr){4-21}
 & & & \rotatebox{90}{$G_{\mathrm{mk}}$} & \rotatebox{90}{$G_{\mathrm{mk}}^{\sigma}$} & \rotatebox{90}{$G_{\mathrm{td}}$} & \rotatebox{90}{$G_{\mathrm{tcd}}$} & \rotatebox{90}{$G_{\mathrm{cj}}$} & \rotatebox{90}{$G_{\mathrm{th}}$} & \rotatebox{90}{$G_{\mathrm{sr}}$} & \rotatebox{90}{$G_{\mathrm{dl}}$} & \rotatebox{90}{$G_{\mathrm{tr}}$} & \rotatebox{90}{$G_{\mathrm{ut}}$} & \rotatebox{90}{$G_{\mathrm{dj}}$} & \rotatebox{90}{$G_{\mathrm{do}}$} & \rotatebox{90}{$G_{\mathrm{en}}$} & \rotatebox{90}{$G_{\mathrm{vc}}$} & \rotatebox{90}{$G_{\mathrm{eo}}$} & \rotatebox{90}{$G_{\mathrm{ed}}$} & \rotatebox{90}{$G_{\mathrm{ds}}$} & \rotatebox{90}{$G_{\mathrm{cm}}$} \\\\
\midrule
$\Gamma^{\mathrm{cen}}$ & LangGraph & GPT-5.4 & 78.3 & 4.59 & 261.2 & 261.2 & 10.0 & 0.128 & 1.00 & 0.000 & 0.000 & 0.000 & 0.000 & 0.000 & 12238.7 & 32.6 & 13454.4 & 9.38 & 120.0 & 120.0 \\
$\Gamma^{\mathrm{cen}}$ & LangGraph & Gemini-3-Flash & 72.9 & 0.961 & 248.4 & 248.4 & 10.0 & 0.137 & 1.00 & 0.000 & 0.000 & 0.000 & 0.000 & 0.000 & 11453.1 & 32.2 & 10924.9 & 4.82 & 120.0 & 120.0 \\
$\Gamma^{\mathrm{cen}}$ & LangGraph & Qwen-3.5 & 74.0 & 1.85 & 253.9 & 253.9 & 10.0 & 0.135 & 1.00 & 0.000 & 0.000 & 0.000 & 0.000 & 0.000 & 11625.5 & 32.0 & 11571.3 & 6.06 & 120.0 & 120.0 \\
$\Gamma^{\mathrm{cen}}$ & AgentScope & GPT-5.4 & 68.6 & 19.7 & 208.3 & 218.5 & 8.30 & 0.112 & 0.800 & 0.000 & 0.200 & 10.1 & 1.70 & 5.30 & 10722.2 & 26.4 & 9912.3 & 7.62 & 106.2 & 105.9 \\
\bottomrule
\end{tabular}%
}
\end{sidewaystable*}

\begin{sidewaystable*}[p]
\centering
\scriptsize
\setlength{\tabcolsep}{1.55pt}
\renewcommand{\arraystretch}{1.05}
\caption[]{Logical full-metric matrix for \texttt{A5C12-2} (Panel B).}
\label{tab:a5c12_cluster_full_b}
\resizebox{\textheight}{!}{%
\begin{tabular}{@{}lll*{20}{c}@{}}
\toprule
\multirow{2}{*}{\textbf{Paradigm}} & \multirow{2}{*}{\textbf{Framework}} & \multirow{2}{*}{\textbf{LLM}} & \multicolumn{20}{c}{\textbf{Panel B: Constraint / Resource / Process}} \\\\
\cmidrule(lr){4-23}
 & & & \rotatebox{90}{$G_{\mathrm{co}}$} & \rotatebox{90}{$G_{\mathrm{vm}}$} & \rotatebox{90}{$G_{\mathrm{com}}$} & \rotatebox{90}{$G_{\mathrm{cod}}$} & \rotatebox{90}{$G_{\mathrm{eof}}$} & \rotatebox{90}{$G_{\mathrm{cof}}$} & \rotatebox{90}{$G_{\mathrm{st}}$} & \rotatebox{90}{$G_{\mathrm{su}}$} & \rotatebox{90}{$G_{\mathrm{tt}}$} & \rotatebox{90}{$G_{\mathrm{te}}$} & \rotatebox{90}{$G_{\mathrm{tm}}$} & \rotatebox{90}{$G_{\mathrm{ic}}$} & \rotatebox{90}{$G_{\mathrm{bt}}$} & \rotatebox{90}{$G_{\mathrm{bw}}$} & \rotatebox{90}{$G_{\mathrm{ae}}$} & \rotatebox{90}{$G_{\mathrm{aj}}$} & \rotatebox{90}{$G_{\mathrm{ab}}$} & \rotatebox{90}{$G_{\mathrm{bd}}$} & \rotatebox{90}{$G_{\mathrm{dt}}$} & \rotatebox{90}{$G_{\mathrm{rt}}$} \\\\
\midrule
$\Gamma^{\mathrm{cen}}$ & LangGraph & GPT-5.4 & 7343.2 & 21527.1 & 8072.7 & 9.38 & 16.3 & 16.3 & 8.94 & 53.6 & 68.6 & 13.7 & 40.0 & 40.0 & 0.000 & 0.000 & 4.00 & 7.00 & 1.75 & 0.300 & 0.136 & 0.136 \\
$\Gamma^{\mathrm{cen}}$ & LangGraph & Gemini-3-Flash & 6871.9 & 17479.8 & 6554.9 & 4.82 & 16.1 & 16.1 & 8.09 & 48.5 & 67.9 & 13.6 & 40.0 & 40.0 & 0.000 & 0.000 & 4.00 & 7.00 & 1.75 & 0.000 & 0.000 & 0.000 \\
$\Gamma^{\mathrm{cen}}$ & LangGraph & Qwen-3.5 & 6975.3 & 18514.0 & 6942.8 & 6.06 & 16.0 & 16.0 & 10.0 & 60.1 & 67.9 & 13.6 & 40.0 & 40.0 & 0.000 & 0.000 & 4.00 & 7.00 & 1.75 & 0.200 & 0.432 & 0.432 \\
$\Gamma^{\mathrm{cen}}$ & AgentScope & GPT-5.4 & 6433.3 & 15859.6 & 5947.4 & 7.62 & 13.2 & 13.2 & 7.41 & 44.4 & 60.8 & 12.2 & 35.5 & 35.5 & 0.000 & 0.000 & 4.00 & 7.00 & 1.75 & 0.100 & 1.06 & 1.06 \\
\bottomrule
\end{tabular}%
}
\end{sidewaystable*}

\begin{sidewaystable*}[p]
\centering
\scriptsize
\setlength{\tabcolsep}{1.7pt}
\renewcommand{\arraystretch}{1.05}
\caption[]{Logical full-metric matrix for \texttt{A5C12-2} (Panel C).}
\label{tab:a5c12_cluster_full_c}
\resizebox{\textheight}{!}{%
\begin{tabular}{@{}lll*{20}{c}@{}}
\toprule
\multirow{2}{*}{\textbf{Paradigm}} & \multirow{2}{*}{\textbf{Framework}} & \multirow{2}{*}{\textbf{LLM}} & \multicolumn{20}{c}{\textbf{Panel C: Coordination / Audit / Diagnostics}} \\\\
\cmidrule(lr){4-23}
 & & & \rotatebox{90}{$G_{\mathrm{vb}}$} & \rotatebox{90}{$G_{\mathrm{bc}}$} & \rotatebox{90}{$G_{\mathrm{pa}}$} & \rotatebox{90}{$G_{\mathrm{ac}}$} & \rotatebox{90}{$G_{\mathrm{dw}}$} & \rotatebox{90}{$G_{\mathrm{ba}}$} & \rotatebox{90}{$G_{\mathrm{pa}}^{\mathrm{amb}}$} & \rotatebox{90}{$G_{\mathrm{aa}}$} & \rotatebox{90}{$G_{\mathrm{ca}}$} & \rotatebox{90}{$G_{\mathrm{as}}$} & \rotatebox{90}{$G_{\mathrm{wc}}$} & \rotatebox{90}{$G_{\mathrm{ar}}$} & \rotatebox{90}{$G_{\mathrm{sv}}$} & \rotatebox{90}{$G_{\mathrm{sc}}$} & \rotatebox{90}{$G_{\mathrm{oc}}$} & \rotatebox{90}{$G_{\mathrm{uc}}$} & \rotatebox{90}{$G_{\mathrm{llm}}$} & \rotatebox{90}{$G_{\mathrm{fb}}$} & \rotatebox{90}{$G_{\mathrm{na}}$} & \rotatebox{90}{$G_{\mathrm{p1}}$} \\\\
\midrule
$\Gamma^{\mathrm{cen}}$ & LangGraph & GPT-5.4 & 2.12 & 40.0 & 0.000 & 0.000 & 2.02 & 0.652 & 0.000 & 0.000 & 0.700 & 120.0 & 155.6 & 120.0 & 160.0 & 10.0 & 30.0 & 0.000 & 120.0 & 0.000 & 0.000 & 1.00 \\
$\Gamma^{\mathrm{cen}}$ & LangGraph & Gemini-3-Flash & 2.00 & 40.0 & 0.000 & 0.000 & 2.02 & 0.560 & 0.000 & 0.000 & 0.700 & 120.0 & 243.8 & 120.0 & 160.0 & 10.0 & 30.0 & 0.000 & 120.0 & 0.000 & 0.000 & -- \\
$\Gamma^{\mathrm{cen}}$ & LangGraph & Qwen-3.5 & 2.04 & 40.0 & 0.000 & 0.000 & 2.02 & 0.602 & 0.000 & 0.000 & 0.698 & 120.0 & 204.5 & 120.0 & 160.0 & 10.0 & 30.0 & 0.000 & 120.0 & 0.000 & 0.000 & 1.00 \\
$\Gamma^{\mathrm{cen}}$ & AgentScope & GPT-5.4 & 2.15 & 35.5 & 0.000 & 0.000 & 2.04 & 0.654 & 0.000 & 0.000 & 0.710 & 106.2 & 354.1 & 106.2 & 140.4 & 8.30 & 27.2 & 0.000 & 106.1 & 0.100 & 0.000 & 0.999 \\
\bottomrule
\end{tabular}%
}
\end{sidewaystable*}

\begin{table*}[htbp]
\centering
\setlength{\tabcolsep}{3pt}
\renewcommand{\arraystretch}{1.08}
\caption{Primary 15-metric matrix for \texttt{A5C12-3}.}
\label{tab:a5c12_late_primary}
\resizebox{\textwidth}{!}{%
\begin{tabular}{lll*{15}{c}}
\toprule
\multirow{3}{*}{\textbf{Paradigm}} & \multirow{3}{*}{\textbf{Framework}} & \multirow{3}{*}{\textbf{LLM}} & \multicolumn{15}{c}{\textbf{Evaluation Metrics}} \\
\cmidrule(lr){4-18}
& & & \multicolumn{3}{c}{$\mathcal{G}_{\mathrm{eff}}$} & \multicolumn{4}{c}{$\mathcal{G}_{\mathrm{con}}$} & \multicolumn{4}{c}{$\mathcal{G}_{\mathrm{coord}}$} & \multicolumn{4}{c}{$\mathcal{G}_{\mathrm{rob}}$} \\
\cmidrule(lr){4-6}\cmidrule(lr){7-10}\cmidrule(lr){11-14}\cmidrule(lr){15-18}
& & & $G_{\mathrm{mk}} \downarrow$ & $G_{\mathrm{td}}\downarrow$ & $G_{\mathrm{cj}}\uparrow$ & $G_{\mathrm{vc}}\downarrow$ & $G_{\mathrm{eo}}\downarrow$ & $G_{\mathrm{ed}}\downarrow$ & $G_{\mathrm{en}}\downarrow$ & $G_{\mathrm{ds}}\downarrow$ & $G_{\mathrm{cm}}\downarrow$ & $G_{\mathrm{fb}}\downarrow$ & $G_{\mathrm{llm}}\downarrow$ & $G_{\mathrm{tr}}\downarrow$ & $G_{\mathrm{dj}}\downarrow$ & $G_{\mathrm{ut}}\downarrow$ & $G_{\mathrm{do}}\downarrow$ \\
\midrule
\multirow{6}{*}{$\Gamma^{\mathrm{cen}}$} & \multirow{3}{*}{LangGraph} & GPT-5.4 & 78.2 & 258.0 & 9.3 & 32.2 & 17323.2 & 9.0 & 12241.0 & 114.0 & 114.0 & 0.1 & 113.9 & 0.10 & 0.7 & 5.2 & 2.3 \\
 &  & Gemini-3-Flash & 79.3 & 265.0 & 10.0 & 34.8 & 14527.0 & 8.3 & 12412.2 & 120.0 & 120.0 & 1.1 & 118.9 & 0.00 & 0.0 & 0.0 & 0.0 \\
 &  & Qwen-3.5 & 80.1 & 272.0 & 10.0 & 34.6 & 15954.8 & 9.2 & 12550.1 & 120.0 & 120.0 & 0.0 & 120.0 & 0.00 & 0.0 & 0.0 & 0.0 \\
\cmidrule(lr){2-18}
 & \multirow{3}{*}{AgentScope} & GPT-5.4 & 83.0 & 284.5 & 10.0 & 35.8 & 18810.1 & 9.9 & 12972.1 & 120.0 & 120.0 & 0.0 & 120.0 & 0.00 & 0.0 & 0.0 & 0.0 \\
 &  & Gemini-3-Flash & 79.3 & 265.0 & 10.0 & 34.8 & 14527.0 & 8.3 & 12412.2 & 120.0 & 120.0 & 1.1 & 118.9 & 0.00 & 0.0 & 0.0 & 0.0 \\
\bottomrule
\end{tabular}%
}
\end{table*}

\begin{sidewaystable*}[p]
\centering
\scriptsize
\setlength{\tabcolsep}{1.9pt}
\renewcommand{\arraystretch}{1.05}
\caption{Logical full-metric matrix for \texttt{A5C12-3} (Panel A).}
\label{tab:a5c12_late_full}
\resizebox{\textheight}{!}{%
\begin{tabular}{@{}lll*{18}{c}@{}}
\toprule
\multirow{2}{*}{\textbf{Paradigm}} & \multirow{2}{*}{\textbf{Framework}} & \multirow{2}{*}{\textbf{LLM}} & \multicolumn{18}{c}{\textbf{Panel A: Primary Outcomes + Robustness Anchors}} \\\\
\cmidrule(lr){4-21}
 & & & \rotatebox{90}{$G_{\mathrm{mk}}$} & \rotatebox{90}{$G_{\mathrm{mk}}^{\sigma}$} & \rotatebox{90}{$G_{\mathrm{td}}$} & \rotatebox{90}{$G_{\mathrm{tcd}}$} & \rotatebox{90}{$G_{\mathrm{cj}}$} & \rotatebox{90}{$G_{\mathrm{th}}$} & \rotatebox{90}{$G_{\mathrm{sr}}$} & \rotatebox{90}{$G_{\mathrm{dl}}$} & \rotatebox{90}{$G_{\mathrm{tr}}$} & \rotatebox{90}{$G_{\mathrm{ut}}$} & \rotatebox{90}{$G_{\mathrm{dj}}$} & \rotatebox{90}{$G_{\mathrm{do}}$} & \rotatebox{90}{$G_{\mathrm{en}}$} & \rotatebox{90}{$G_{\mathrm{vc}}$} & \rotatebox{90}{$G_{\mathrm{eo}}$} & \rotatebox{90}{$G_{\mathrm{ed}}$} & \rotatebox{90}{$G_{\mathrm{ds}}$} & \rotatebox{90}{$G_{\mathrm{cm}}$} \\\\
\midrule
$\Gamma^{\mathrm{cen}}$ & LangGraph & GPT-5.4 & 78.2 & 15.0 & 258.0 & 263.2 & 9.30 & 0.117 & 0.900 & 0.000 & 0.100 & 5.21 & 0.700 & 2.30 & 12241.0 & 32.2 & 17323.2 & 9.04 & 114.0 & 114.0 \\
$\Gamma^{\mathrm{cen}}$ & LangGraph & Gemini-3-Flash & 79.3 & 4.78 & 265.0 & 265.0 & 10.0 & 0.127 & 1.00 & 0.000 & 0.000 & 0.000 & 0.000 & 0.000 & 12412.2 & 34.8 & 14527.0 & 8.28 & 120.0 & 120.0 \\
$\Gamma^{\mathrm{cen}}$ & LangGraph & Qwen-3.5 & 80.1 & 4.20 & 272.0 & 272.0 & 10.0 & 0.125 & 1.00 & 0.000 & 0.000 & 0.000 & 0.000 & 0.000 & 12550.1 & 34.6 & 15954.8 & 9.16 & 120.0 & 120.0 \\
$\Gamma^{\mathrm{cen}}$ & AgentScope & GPT-5.4 & 83.0 & 3.77 & 284.5 & 284.5 & 10.0 & 0.121 & 1.00 & 0.000 & 0.000 & 0.000 & 0.000 & 0.000 & 12972.1 & 35.8 & 18810.1 & 9.88 & 120.0 & 120.0 \\
\bottomrule
\end{tabular}%
}
\end{sidewaystable*}

\begin{sidewaystable*}[p]
\centering
\scriptsize
\setlength{\tabcolsep}{1.55pt}
\renewcommand{\arraystretch}{1.05}
\caption[]{Logical full-metric matrix for \texttt{A5C12-3} (Panel B).}
\label{tab:a5c12_late_full_b}
\resizebox{\textheight}{!}{%
\begin{tabular}{@{}lll*{20}{c}@{}}
\toprule
\multirow{2}{*}{\textbf{Paradigm}} & \multirow{2}{*}{\textbf{Framework}} & \multirow{2}{*}{\textbf{LLM}} & \multicolumn{20}{c}{\textbf{Panel B: Constraint / Resource / Process}} \\\\
\cmidrule(lr){4-23}
 & & & \rotatebox{90}{$G_{\mathrm{co}}$} & \rotatebox{90}{$G_{\mathrm{vm}}$} & \rotatebox{90}{$G_{\mathrm{com}}$} & \rotatebox{90}{$G_{\mathrm{cod}}$} & \rotatebox{90}{$G_{\mathrm{eof}}$} & \rotatebox{90}{$G_{\mathrm{cof}}$} & \rotatebox{90}{$G_{\mathrm{st}}$} & \rotatebox{90}{$G_{\mathrm{su}}$} & \rotatebox{90}{$G_{\mathrm{tt}}$} & \rotatebox{90}{$G_{\mathrm{te}}$} & \rotatebox{90}{$G_{\mathrm{tm}}$} & \rotatebox{90}{$G_{\mathrm{ic}}$} & \rotatebox{90}{$G_{\mathrm{bt}}$} & \rotatebox{90}{$G_{\mathrm{bw}}$} & \rotatebox{90}{$G_{\mathrm{ae}}$} & \rotatebox{90}{$G_{\mathrm{aj}}$} & \rotatebox{90}{$G_{\mathrm{ab}}$} & \rotatebox{90}{$G_{\mathrm{bd}}$} & \rotatebox{90}{$G_{\mathrm{dt}}$} & \rotatebox{90}{$G_{\mathrm{rt}}$} \\\\
\midrule
$\Gamma^{\mathrm{cen}}$ & LangGraph & GPT-5.4 & 7344.6 & 27717.1 & 10393.9 & 9.04 & 16.1 & 16.1 & 8.90 & 53.4 & 68.6 & 13.7 & 38.1 & 38.1 & 0.000 & 0.000 & 4.00 & 7.00 & 1.75 & 1.00 & 2.26 & 2.26 \\
$\Gamma^{\mathrm{cen}}$ & LangGraph & Gemini-3-Flash & 7447.3 & 23243.1 & 8716.2 & 8.28 & 17.4 & 17.4 & 8.68 & 52.1 & 70.1 & 14.0 & 40.0 & 40.0 & 0.000 & 0.000 & 4.00 & 7.00 & 1.75 & 0.500 & 2.62 & 2.62 \\
$\Gamma^{\mathrm{cen}}$ & LangGraph & Qwen-3.5 & 7530.1 & 25527.6 & 9572.9 & 9.16 & 17.3 & 17.3 & 8.85 & 53.1 & 70.3 & 14.1 & 40.0 & 40.0 & 0.000 & 0.000 & 4.00 & 7.00 & 1.75 & 0.500 & 2.37 & 2.37 \\
$\Gamma^{\mathrm{cen}}$ & AgentScope & GPT-5.4 & 7783.3 & 30096.2 & 11286.1 & 9.88 & 17.9 & 17.9 & 9.30 & 55.8 & 72.0 & 14.4 & 40.0 & 40.0 & 0.000 & 0.000 & 4.00 & 7.00 & 1.75 & 0.500 & 1.75 & 1.75 \\
\bottomrule
\end{tabular}%
}
\end{sidewaystable*}

\begin{sidewaystable*}[p]
\centering
\scriptsize
\setlength{\tabcolsep}{1.7pt}
\renewcommand{\arraystretch}{1.05}
\caption[]{Logical full-metric matrix for \texttt{A5C12-3} (Panel C)}
\label{tab:a5c12_late_full_c}
\resizebox{\textheight}{!}{%
\begin{tabular}{@{}lll*{20}{c}@{}}
\toprule
\multirow{2}{*}{\textbf{Paradigm}} & \multirow{2}{*}{\textbf{Framework}} & \multirow{2}{*}{\textbf{LLM}} & \multicolumn{20}{c}{\textbf{Panel C: Coordination / Audit / Diagnostics}} \\\\
\cmidrule(lr){4-23}
 & & & \rotatebox{90}{$G_{\mathrm{vb}}$} & \rotatebox{90}{$G_{\mathrm{bc}}$} & \rotatebox{90}{$G_{\mathrm{pa}}$} & \rotatebox{90}{$G_{\mathrm{ac}}$} & \rotatebox{90}{$G_{\mathrm{dw}}$} & \rotatebox{90}{$G_{\mathrm{ba}}$} & \rotatebox{90}{$G_{\mathrm{pa}}^{\mathrm{amb}}$} & \rotatebox{90}{$G_{\mathrm{aa}}$} & \rotatebox{90}{$G_{\mathrm{ca}}$} & \rotatebox{90}{$G_{\mathrm{as}}$} & \rotatebox{90}{$G_{\mathrm{wc}}$} & \rotatebox{90}{$G_{\mathrm{ar}}$} & \rotatebox{90}{$G_{\mathrm{sv}}$} & \rotatebox{90}{$G_{\mathrm{sc}}$} & \rotatebox{90}{$G_{\mathrm{oc}}$} & \rotatebox{90}{$G_{\mathrm{uc}}$} & \rotatebox{90}{$G_{\mathrm{llm}}$} & \rotatebox{90}{$G_{\mathrm{fb}}$} & \rotatebox{90}{$G_{\mathrm{na}}$} & \rotatebox{90}{$G_{\mathrm{p1}}$} \\\\
\midrule
$\Gamma^{\mathrm{cen}}$ & LangGraph & GPT-5.4 & 2.07 & 38.1 & 0.000 & 0.000 & 2.02 & 0.601 & 0.000 & 0.000 & 0.702 & 114.0 & 158.1 & 114.0 & 151.6 & 9.30 & 28.8 & 0.000 & 113.9 & 0.100 & 0.000 & 0.999 \\
$\Gamma^{\mathrm{cen}}$ & LangGraph & Gemini-3-Flash & 1.97 & 40.0 & 0.000 & 0.000 & 2.02 & 0.530 & 0.000 & 0.000 & 0.700 & 120.0 & 250.7 & 120.0 & 160.0 & 10.0 & 30.0 & 0.000 & 118.9 & 1.10 & 0.000 & -- \\
$\Gamma^{\mathrm{cen}}$ & LangGraph & Qwen-3.5 & 1.97 & 40.0 & 0.000 & 0.000 & 2.02 & 0.565 & 0.000 & 0.000 & 0.700 & 120.0 & 201.2 & 120.0 & 160.0 & 10.0 & 30.0 & 0.000 & 120.0 & 0.000 & 0.000 & 1.00 \\
$\Gamma^{\mathrm{cen}}$ & AgentScope & GPT-5.4 & 2.01 & 40.0 & 0.000 & 0.000 & 2.02 & 0.570 & 0.000 & 0.000 & 0.700 & 120.0 & 294.7 & 120.0 & 160.0 & 10.0 & 30.0 & 0.000 & 120.0 & 0.000 & 0.000 & 1.00 \\
\bottomrule
\end{tabular}%
}
\end{sidewaystable*}

%
\section{Using the Benchmark} \label{ues}
DESBench is designed for researchers with existing MAS implementations or those developing new coordination methods, who wish to evaluate coordination effectiveness under shared event-driven scheduling dynamics. The benchmark provides the instance configurations, simulator, legal-action validation, execution traces, and evaluation metrics, while the user supplies a MAS runner that consumes the public decision payload and returns valid actions. The recommended usage follows a simple pipeline: select a reported instance, connect the runner, optionally register a new coordination paradigm, execute the provided entry points, and inspect the resulting summaries and trace artifacts.

\subsection{Choose a Reported Instance}
A valid comparison begins by fixing one reported instance in Table~\ref{tab:appendix_instance_catalog}. The reported instance already fixes topology, route grammar, arrival process, disturbance profile, budgets, and seed split. The four paper instances cover distinct coordination stresses: \texttt{A3C9-1} for smaller-topology transfer with cross-area coupling, \texttt{A5C12-1} for branch pressure and routing ambiguity, \texttt{A5C12-2} for cluster pull and greedy-routing lure, and \texttt{A5C12-3} for late commitment and long-horizon regret. Once the instance is chosen, it should remain fixed for all methods in the same comparison.

\subsection{Connect a MAS Runner}

The main integration point is the MAS runner. At each decision step, DESBench provides a decision payload containing the active agents and their local views. Each view includes the agent role, observation, protocol-visible context, and an action mask. The runner must return exactly one valid action for each active agent, consistent with the provided action mask.

\begin{nolinenumbers}
\begin{lstlisting}[style=promptblock]
class MyMASRunner:
    def select_actions(self, decision):
        actions = {}
        for agent_id, view in decision["active_agents"].items():
            mask = view["action_mask"]
            actions[agent_id] = choose_legal_action(view, mask)
        return actions
\end{lstlisting}
\end{nolinenumbers}

This interface is framework-agnostic. The runner can be implemented using LLMs, learned policies, rule-based logic, or custom orchestration. It operates solely on the public payload and must respect the active-agent set and action constraints.

\subsection{Register a New Paradigm}
If the contribution is a new coordination paradigm, it should be assigned a public id and registered before running the suite. A paradigm registration specifies the benchmark-visible coordination semantics, including routing authority, rejection rights, local rerouting, escalation conditions, and negotiation mechanisms. The repository provides \path{custom_paradigm.py} as a minimal template.

\begin{nolinenumbers}
\begin{lstlisting}[style=promptblock]
from fms_bench.paradigms import AuthorityMode, register_authority_mode

register_authority_mode(
    AuthorityMode(
        mode_id="my_hybrid",
        routing_authority_mode="holonic_hybrid",
        cell_contract_actions="accept_reject",
        area_local_reroute_enabled=True,
        plant_reroute_required=False,
        mediated_bid_round_enabled=True,
    )
)
\end{lstlisting}
\end{nolinenumbers}



This registration defines the coordination semantics exposed by DESBench, while the MAS runner determines how agents act under these semantics. The coordination structure may be static or dynamic. For example, authority and interaction patterns may shift over time between local and global coordination regimes depending on system state or event context.

If the paradigm can be expressed through existing authority switches, a lightweight registration is sufficient. More substantial extensions (e.g., new activation types, message objects, legal actions, or settlement semantics) require corresponding runtime modifications and validation.

For paradigm-level comparison, the reported instance, seed split, runner interface, simulator dynamics, and metrics should remain fixed. The paradigm may change authority, information flow, negotiation, rerouting, or escalation, but should not alter topology, route grammar, hidden simulator state, or metric definitions.

\subsection{Run and Audit}
DESBench separates MAS connection from experiment execution. For an existing external MAS, \path{run_mas.py} is the recommended adapter because it exercises the same public decision payload without requiring the MAS to be written in the repository. For built-in or repository-native runners, \path{run_suite.py} executes a full suite and produces a standard \path{suite_summary.csv}, while \path{run_case.py} reruns selected reported cases for detailed diagnosis or reproduction. For quick testing, users may restrict cases and seeds and use a lightweight backend. For comparison, all debug limits should be removed, and the reported instance, paradigm id, runner, model setting, and seed split must be kept identical across methods within a comparison block. Examples are shown below.
\begin{table*}[htbp]
\centering
\scriptsize
\caption{Linear use checklist for connecting a MAS design to DESBench.}
\label{tab:appendix_run_configuration_checklist}
\setlength{\tabcolsep}{4pt}
\begin{tabular}{@{}L{0.18\linewidth} L{0.36\linewidth} L{0.38\linewidth}@{}}
\toprule
\textbf{Step} & \textbf{What DESBench fixes} & \textbf{What the user provides} \\
\midrule
Choose instance
& Reported instance, topology, route grammar, arrivals, disturbances, budgets, and metrics.
& Select \texttt{A3C9-1}, \texttt{A5C12-1}, \texttt{A5C12-2}, or \texttt{A5C12-3}. \\
Connect runner
& Active-agent payloads, local observations, protocol context, and legal-action masks.
& Return one legal action for each active agent through \path{select_actions} or the \path{run_mas.py} JSON boundary. \\
Declare paradigm
& Built-in protocol semantics for \path{centralized}, \path{hierarchical}, \path{heterarchical_cnp}, \path{holonic_hybrid}, or newly proposed coordination paradigms.
& Reuse a built-in paradigm or register a new \path{mode_id} for the proposed coordination semantics. \\
Run comparison
& Public runners, locked test seeds, output schema, and trace format.
& Use the same instance, seed split, runner boundary, model setting, and metrics for all compared methods. \\
Audit result
& Metric summaries, runtime traces, and decision-audit fields.
& Check coverage, fallback behavior, invalid actions, no-action epochs, and decision failures. \\
\bottomrule
\end{tabular}
\setlength{\tabcolsep}{6pt}
\end{table*}

\begin{nolinenumbers}
\begin{lstlisting}[style=promptblock]
python examples/run_mas.py \\
  --suite intercell_a3c9_wide \\
  --authority-mode heterarchical_cnp \\
  --output-dir dist/benchmark_runs/my_mas_a3c9 \\
  --controller-command python my_mas_runner.py
\end{lstlisting}

\begin{lstlisting}[style=promptblock]
python examples/run_suite.py \\
  --framework langgraph \\
  --suite intercell_a3c9_wide \\
  --authority-mode heterarchical_cnp \\
  --llm-backend openai_compat \\
  --llm-model qwen/qwen3.5-plus-02-15 \\
  --output-dir dist/benchmark_runs/a3c9_cross_langgraph_heter
\end{lstlisting}

\begin{lstlisting}[style=promptblock]
python examples/run_case.py \\
  --framework agentscope \\
  --suite intercell_a5c12_harder \\
  --case-id <exact_case_id_from_suite_summary_or_artifacts> \\
  --authority-mode holonic_hybrid \\
  --llm-backend openai_compat \\
  --llm-model openai/gpt-5.4 \\
  --output-dir dist/benchmark_runs/a5c12_case_shards
\end{lstlisting}
\end{nolinenumbers}

\subsection{Read the Comparison Surface}
The resulting runs should be inspected first through the 15 primary metrics in Tables~\ref{tab:a3c9_cross_primary}-\ref{tab:a5c12_late_full_c}, which form the main comparison surface. These metrics cover effectiveness, constraint alignment, coordination efficiency, and robustness, and no single scalar score is sufficient to establish superiority. Each run also produces a standardized artifact set (Table~\ref{tab:appendix_output_artifacts}) to support reconstruction and diagnosis. We encourage users to design adaptive coordination strategies that dynamically adjust between local and global behaviors over time, aiming to achieve strong performance across most metric families.

\begin{table*}[htbp]
\centering
\scriptsize
\caption{Expected DESBench output artifacts for user-side evaluation and diagnosis.}
\label{tab:appendix_output_artifacts}
\setlength{\tabcolsep}{4pt}
\begin{tabular}{@{}L{0.20\linewidth} L{0.34\linewidth} L{0.38\linewidth}@{}}
\toprule
\textbf{Artifact type} & \textbf{Representative files or fields} & \textbf{How users should use them} \\
\midrule
Metrics summary
& \path{suite_summary.csv} and per-case \path{summary.csv}
& Start here for the primary 15 metrics and the broader Appendix~\ref{app:metric_catalog} surface. Use these files to compare methods on the same reported instance. \\
Artifact manifest
& Per-case \path{artifacts.csv}
& Use it to locate the exact trace files and verify that the reported instance, runner, backend, and budget metadata match the intended experiment. \\
World and activation traces
& \path{*_world_event_trace.json} and \path{*_activation_trace.json}
& Diagnose when queueing, routing, transport, breakdowns, or activation patterns cause a coordination method to slow down or deadlock. \\
Settlement and protocol traces
& \path{*_settlement_trace.json} and \path{*_protocol_trace.json}
& Diagnose contract settlement, reroute, rejection, escalation, and message-level coordination behavior induced by $\Gamma$. \\
Decision audit fields
& \path{framework_llm_decision_count}, \path{framework_fallback_contract_decision_count}, \path{framework_no_legal_action_count}, and related traces
& Distinguish model-backed decisions from fallback and no-legal-action cases before concluding that one method is genuinely stronger. \\
Failure and fallback logs
& \path{progress_log.jsonl}, \path{progress_snapshot.json}, and failure traces
& Check exhausted retries, invalid responses, interrupted cases, and other execution failures that can confound comparison if left unexplained. \\
\bottomrule
\end{tabular}
\setlength{\tabcolsep}{6pt}
\end{table*}


\clearpage

\end{document}